\documentclass[useAMS,usenatbib]{mn2e}

  \usepackage{graphicx}
   \usepackage{lscape}
 \usepackage{rotating}
  \usepackage{array}
  \usepackage{flafter}
  \usepackage{caption}
  \usepackage{subfigure}

\title{ The bulge-disk decomposition of AGN host galaxies.}
\author[V. A. Bruce]
{V.\,A. Bruce$^{1}$\thanks{E-mail: vab@roe.ac.uk}, J.\,S. Dunlop$^{1}$, A. Mortlock$^{1}$, D.\,D. Kocevski$^{2}$, E.\,J. McGrath$^{2}$, 
\newauthor D.\,J. Rosario$^{3}$\\
$^1$SUPA\thanks{Scottish Universities Physics Alliance} Institute for Astronomy, University of Edinburgh, Royal Observatory, Edinburgh EH9 3HJ\\
$^2$Department of Physics and Astronomy, Colby College, Waterville, ME 04901, USA\\
$^3$ Max-Planck-Institut fu\"r Extraterrestrische Physik (MPE), Postfach 1312, 85741 Garching, Germany}
\begin{document}

\date{}

\pagerange{\pageref{firstpage}--\pageref{lastpage}} \pubyear{2015}

\pagestyle{myheadings}
\markboth{V. A. Bruce et al.} {Decomposed morphologies of AGN hosts at $0.5 < z < 3$}

\maketitle

\label{firstpage}

\begin{abstract}
 We present the results from a study of the morphologies of moderate luminosity X-ray selected AGN host galaxies in comparison to a carefully mass-matched control sample at $0.5 < z < 3$ in the CANDELS GOODS-S field. We apply a multi-wavelength morphological decomposition analysis to these two samples and report on the differences between the morphologies as fitted from single S\'{e}rsic and multiple S\'{e}rsic models, and models which include an additional nuclear point-source component. Thus, we are able to compare the widely adopted single S\'{e}rsic fits from previous studies to the results from a full morphological decomposition, and address the issue of how biased the inferred properties of AGN hosts are by a potential nuclear contribution from the AGN itself. We find that the AGN hosts are mixed systems which have higher bulge fractions than the control sample in our highest redshift bins at the $>99.7 \%$ confidence level, according to all model fits even those which adopt a point-source component. This serves to alleviate concerns that previous, purely single S\'{e}rsic, analyses of AGN hosts could have been spuriously biased towards higher bulge fractions.
This dataset allows us to further probe the physical nature of these point-source components; we find no strong correlation between the point-source component and AGN activity, and that these point-source components are best modelled physically by nuclear starbursts. Our analysis of the bulge and disk fractions of these AGN hosts in comparison to a mass-matched control sample reveals a similar morphological evolutionary track for both the active and non-active populations, providing further evidence in favour of a model where AGN activity is triggered by secular processes.

\end{abstract}


\section{Introduction}

The physical processes responsible for triggering active galactic nuclei (AGN) are still uncertain. Despite the fact that many galaxy evolution models invoke major mergers to both trigger AGN and transform the underlying host galaxy morphologies from disk to bulge dominated systems (e.g. \citealt{Hopkins2005, Springel2005c}), there is mounting observational evidence that mergers may not play a dominant role and instead AGN activity may be driven by more secular processes \citep{Gabor2009,Cisternas2011,Kocevski2012,Schawinski2012,Rosario2015}

If major mergers are the main mechanism for fuelling AGN activity, then the associated morphological changes of the host galaxies should provide clear evidence of this interaction. This evidence could manifest as a change from disk to bulge-dominated systems, or, if observed during the transformation period, to mixed bulge+disk systems exhibiting highly disturbed morphologies.

Locally, AGN hosts are indeed found to reside in more bulge-dominated systems (e.g. \citealt{Kauffmann2003b}) and there are several well-known relations such as the strong correlation between black hole mass and bulge luminosity (e.g.\citealt{Kormendy1995,Magorrian1998}) and the link between black hole mass and stellar velocity dispersion (e.g. \citealt{Gebhardt2000,Ferrarese2000}) that suggest there is a strong connection between, and potentially co-evolution of, the central black hole and the galactic bulge.

At higher redshifts  ($z\geq 1$) we also see evidence of AGN hosts displaying more bulge dominated, or mixed bulge+disk morphologies from high resolution studies conducted with ACS on {\it HST}. From comparing the hosts of X-ray selected moderate luminosity AGN to a control sample, \citet{Grogin2005} showed through the use of non-parametric concentration measurements that the AGN hosts are more bulge dominated. Moreover, the study of \citet{Silverman2008} reported that AGN reside predominantly in the ``green valley", often thought of as a transition region in colour-magnitude parameter space, and consistent with the picture of merger induced AGN activity and morphological transformations. 

However, there is also growing evidence within this higher redshift regime that AGN hosts do not display any excess of disturbed morphologies, another key signature of mergers, in comparison to control samples. Using both visual classifications and parametric fits of a sample of  $z\sim1$ X-ray selected AGN, \citet{Gabor2009} and \citet{Cisternas2011} observe that the host galaxies are no more disturbed than a control sample. Overall the AGN host morphologies in these studies do appear to be transitioning from disk to bulge-dominated but as many of the AGN hosts retain disks, the authors conclude that the AGN activity in these systems has not been triggered by major mergers.
In fact, \citet{Cisternas2011} note that the lack of morphological signatures of major mergers, in terms of disturbed structures, cannot simply be explained by a time lag between the merger event and the time at observation, because if these systems had undergone a merger prior to their epoch of observation one would not expect to find the AGN in such dynamically relaxed disk systems.

With the advent of high resolution near-infrared imaging provided by WFC3/IR on {\it HST,} the CANDELS survey has allowed the rest-frame optical morphologies of AGN hosts in this redshift regime to be explored in more detail. \citet{Kocevski2012} and \citet{Rosario2015} confirm that while the majority of X-ray selected AGN hosts at $z\sim2$ are disks, in comparison to a mass-matched non-active control sample, the AGN hosts have higher bulge fractions. However, these studies have been limited to single S\'{e}rsic morphological fits and/or visual classifications and so do not directly trace the different bulge and disk fractions within the two populations. 

\citet{Schawinski2012} extended the rest-frame optical analysis of AGN hosts to include a heavily obscured AGN sample. As simulations predict that major mergers will first trigger obscured AGN systems, which then become unobscured at later times, this analysis searched for morphological features of a merger in less evolved AGN systems. \citet{Schawinski2012} found no evidence for disturbed morphologies within this sample and concluded that the majority of heavily obscured AGN also resided in disk-dominated hosts (again using single S\'{e}rsic classifications, albeit with careful examination of residuals for additional structure), seeming to confirm the growing picture that AGN activity is not fuelled by mergers, but instead may be triggered by secular processes. However, see also \citet{Kocevski2015} for a similar study of spectroscopically identified Compton-thick AGN.

This observed trend for AGN hosts to exhibit disk-dominated morphologies, with larger contributions from a bulge component than matched control samples, makes it clear that both the bulge and disk components play an important role in the triggering of AGN. Further insight into this aspect of AGN evolution is given by \citet{Schramm2013} who decompose AGN host morphologies and compare the $M_{BH}-M_{total}$ and $M_{BH}-M_{bulge}$ relations at z$\sim1$ to their local counterparts. They find that, while the total stellar mass in the high-redshift AGN hosts is already in agreement with the local relation, there must be a transfer of mass from the disk to the bulge in order to build up the local $M_{BH}-M_{bulge}$  relation since z$\sim1$. It is not clear that mergers play the dominant role in this process and in fact secular processes such as violent disk instabilities (e.g. \citealt{Dekel2009a}; \citealt{Ceverino2010}) may be key for such a transformation, as they are thought to be in the evolution of non-active massive galaxies within this redshift regime.
These findings again lend support to a scenario for AGN hosts in which a viable explanation for their mixed bulge+disk morphologies and a self-consistent build up towards local relations is not (at least solely) dependent on major mergers.

In order to better explore the influence of both the bulge and disk components on the evolution of the AGN, in this paper we conduct a full multi-wavelength decomposition of AGN hosts and a large mass-matched control sample into the separate bulge and disk components and test how the different morphological analyses often adopted in literature effect the best-fit morphologies of the hosts. 
This analysis utilises the morphological decomposition technique presented in \citet{Bruce2012}, which was extended to encompass full multi-wavelength decompositions in \citet{Bruce2014} and applied to samples in both the CANDELS UDS and COSMOS fields for $M_*>10^{11}\,{\rm M_{\odot}}$ galaxies at $1<z<3$.  For the purposes of this paper we concentrate on the $H_{160}$-band morphological decompositions results, adopting light fractions for the individual components measured from the $H_{160}$-band decompositions, and quoting decomposed stellar masses for the bulge and disk components based on the total stellar masses of the systems subdivided according to the light fractions for each component (as was shown in \citealt{Bruce2014} to be representative of the fully decomposed SED fitted stellar masses). A full multi-wavelength decomposition using the accompanying WFC3 and ACS CANDELS data will be presented in a companion paper (Bruce et al., in preparation), where we extend our multiple component + point-source decompositions to the shorter wavebands and conduct bulge+disk+point-source SED fitting where motivated. This multi-wavelength analysis has also allowed us to further address the physical nature of the morphologically identified point-source components in both the AGN hosts and the control sample galaxies, via template fitting of the fully decomposed point-source SEDs, and we conclude in our companion paper that the majority of these point-source components are in fact nuclear starbursts, as will be discussed in Section 5.

This is the first work to self-consistently decompose the morphologies for a large sample of mass-matched control objects and AGN hosts at $z>0.5$ and as such provides new insight into the directly measured bulge and disk fractions of the active and non-active populations. We also address how the adoption of a nuclear point source impacts the host fits, as there is some suggestion from previous studies that the more bulge-dominated morphologies fitted to AGN hosts may be biased by a contribution from the AGN itself. 

From the alternative perspective, point-source fits are also adopted in non-active galaxy fits, often in single S\'{e}rsic fits which would otherwise exceed $n=10$, and are deemed to be motivated by the presence of either an AGN or a nuclear starburst. Thus, this work provides the ideal dataset with which to test the physical nature of the point-source components in addition to exploring how they impact the fits of known AGN.

This paper is structured as follows. In Section 2 we describe the dataset used for this work, discuss the photometric redshift and stellar-mass fitting employed and present our mass-matching technique. Following this, in Section 3 we detail the morphology fitting techniques adopted. In Section 4 we present the results of our analysis for both the single and multiple component S\'{e}rsic models and discuss in detail the impact of the addition of point-source components on the fitted morphologies. In this section we also briefly explore the connection between black hole (BH) mass and both total and bulge stellar mass by investigating trends within our sample selection. Finally, in Section 5 we  extend this discussion to the physical nature of these point-source components by exploring their trends with various observed properties. We conclude in Section 6 with a summary of our findings placed within the context of current models for AGN drivers and galaxy evolution as a whole.

Throughout we calculate all physical
quantities assuming a $\Lambda$CDM universe with $\Omega_{m}=0.3$, $\Omega_{\Lambda}=0.7$ and $H_{0}=70\rm{kms^{-1}Mpc^{-1}}$.

\section{Data}
For this work we have used the HST WFC3/IR and accompanying multi-wavelength data from the CANDELS multi-cycle treasury programme \citep{Grogin2011, Koekemoer2011} in the GOODS-S field. The CANDELS GOODS-S field has been observed as part of the various tiers of the CANDELS observing strategy and so comprises CANDELS  wide+deep+ERS imaging and covers a total area of $\sim170$ square arcmins in the F160W and F125W filters, with the deep and wide fields covered in the F105W filter and the ERS field covered in the F098M filter.
The corresponding 5-$\sigma$ point-source depths in the F160W filters are 27.6 (AB mag) in deep, 26.8 in wide and 27.3 in the ERS field. 
The CANDELS imaging in the GOODS-S field is accompanied by extensive existing multi-wavelength data ranging from the X-ray to the mid-infrared and UV. For the purposes of this work we make use of the optical imaging from HST/ACS in the F850LP, F775W, F606W and F435W  taken as part of the GOODS Hubble Treasury Program \citep{Giavalisco2004}, and combine this with:  U-band from VIMOS \citep{Nonino2009} and CTIO; K$_{s}$-band from ISAAC \citep{Retzlaff2010}; and Spitzer/IRAC $3.6-8\mu$m imaging from the GOODS Spitzer Legacy Programme (PI Dickinson) and SEDS (PI Fazio) \citep{Ashby2013}.

In order to make use of the high resolution HST/WFC3 and ACS imaging in combination with the low-resolution ground-based and confused mid-IR data, we adopt the GOODS-S CANDELS catalogue of \citet{Guo2013}, which is generated from the co-added max-depth F160W GOODS-S mosaic for the deep+wide+ERS fields. In brief, the \citet{Guo2013} catalogue is constructed from an adapted version of {\sc sextractor} \citep{Galametz2013} run on hot and cold mode on the F160W mosaic and merged \citep{Barden2012} to produce a detection catalogue, which is then used in {\sc sextractor} dual mode on the PSF matched HST/WFC3 and ACS mosaics to provide a multi-wavelength catalogue. The flux measurements for the objects in this catalogue are corrected to total based on an aperture correction determined in the F160W detection-band of  $\frac{flux_{auto}}{flux_{iso}}$, for all but the smallest objects (isophotal radii$<2$ pixels) for which the accuracy of the PSF matching becomes significant and for which the aperture correction is given by $\frac{flux_{auto}}{flux_{aper, 2pixels}}$. The total fluxes from the lower-resolution and confused data are then measured using the TFIT code \citep{Laidler2007} which is based on a template fitting approach. It is the \citet{Guo2013} photometry catalogue which has been used for photometric redshift fitting and which we use for stellar-mass fitting.

\subsection{Photometric Redshift and Stellar-Mass Fitting}
The photometric redshifts used in this work are taken from the \citet{Dahlen2013} GOODS-S CANDELS catalogue. The \citet{Dahlen2013} analysis explores the effects on redshift fitting from different codes, stellar population templates and the implementation of additional fitting procedures such as fitting emission lines, adopting photometric zero-point corrections from a spectroscopic training set and interpolating between templates. \citeauthor{Dahlen2013} explore the redshift fits from 11 different groups and compare to a spectroscopic control sample of $\sim600$ galaxies. All of the fitting codes use the same photometric catalogue taken from  \citet{Guo2013}. The implemented catalogue contains 14 bands for each object: U-band VIMOS, B,V,i,z-band HST/ACS, F098M or F105W, F125W, F160W HST/WFC3, K$_{s}$ ISAAC, 3.6, 4.5, 5.8, 8$\mu$m Spitzer/IRAC.
Overall they do not find that a specific code performs significantly better than the others but that the scatter in, and the outlier fraction of, the best-fit photometric redshifts is best improved by combining the multiple fits, although the exact procedure for combining the results does not strongly influence the best-fits. Here we use the best-fit photometric redshifts from 6 of the fitting codes combined with a hierarchical Bayesian method  \citep{Lang2012}.

We evaluate stellar masses using the best-fit photometric redshifts from \citet{Dahlen2013}. These masses have been estimated using the SED fitting code based on HYPERZ \citep{Bolzonella2000,Cirasuolo2007,Bruce2012} with a 12-band photometry catalogue taken from \citet{Guo2013}: U CTIO and VIMOS, B,V,i,z-band HST/ACS, F098M (ERS field) or F105 (deep+wide fields), F125W, F160W HST/WFC3, K$_{s}$ ISAAC, 3.6 and 4.5$\mu$m Spitzer/IRAC. For the fitting we adopt the Bruzual \& Charlot (2003) models with single-component exponentially decaying star-formation histories with e-folding times in the range $0.3< \tau({\rm Gyr})< 5$, with a minimum model age limit of 50 Myr and a Chabrier IMF. Absorption from the intergalactic medium is accounted for using the prescriptions of \citet{Madau1995}, and the \citet{Calzetti2000} obscuration law is used to account for reddening due to dust within the range $0\leq Av \leq 4$.

\subsection{Sample Selection}
Given that the aim of this work is to explore the decomposed morphologies of AGN hosts within the context of a matched sample of non-active galaxies, we have utilised the \citet{Hsu2014} counterparts of the \citet{Xue2011} and \citet{Rangel2014} 4Ms Chandra catalogue to select moderate luminosity, $L_{X}>1\times10^{41} {\rm erg\,s^{-1}}$, AGN from which we define a master sample of galaxies containing these AGN candidates (which have a median X-ray luminosity of $4.9\times10^{42} {\rm erg\,s^{-1}}$). In order to maximise the overlap between our samples (as discussed further in Section 4.4), we adopt a redshift range of $0.5<z<3$ and a stellar mass threshold of $M_*>10^{10}\,{\rm M_{\odot}}$. Within this sample 90\% of objects are $\ga$ 45$\sigma$ detections in the $H_{160}$-band, providing acceptable signal-to-noise for the morphological decompositions.

\subsection{Mass and Redshift Matching}
For a sample cut at $M_*>10^{10}\,{\rm M_{\odot}}$ the mass distribution for AGN host galaxies is centred towards higher masses than that of the control sample of galaxies \citep{Rosario2013}. As a result, in order to conduct a robust comparison between the physical properties of these two populations they must be matched in mass (within redshift bins of $\Delta z=0.5$ to negate any differences in mass-redshift trends), so as not to introduce any biases. However, in order to fully exploit our much larger non-active galaxy sample ($\sim$1600 objects ), for each redshift bin we create a 1000 times bootstrapped, mass-matched, control sample similar to the procedure adopted in \citet{Rosario2015}, but crucially containing the maximal number of control objects as is allowed by the data whilst still ensuring that the mass distributions in the control and AGN host samples are consistent at the $>95\%$ confidence level using a two sample Kolmogorov-Smirnov (K-S) test.

 \begin{figure}
\begin{center} 
 \includegraphics[scale=0.5, trim=1.5cm 13cm 0cm 3cm] {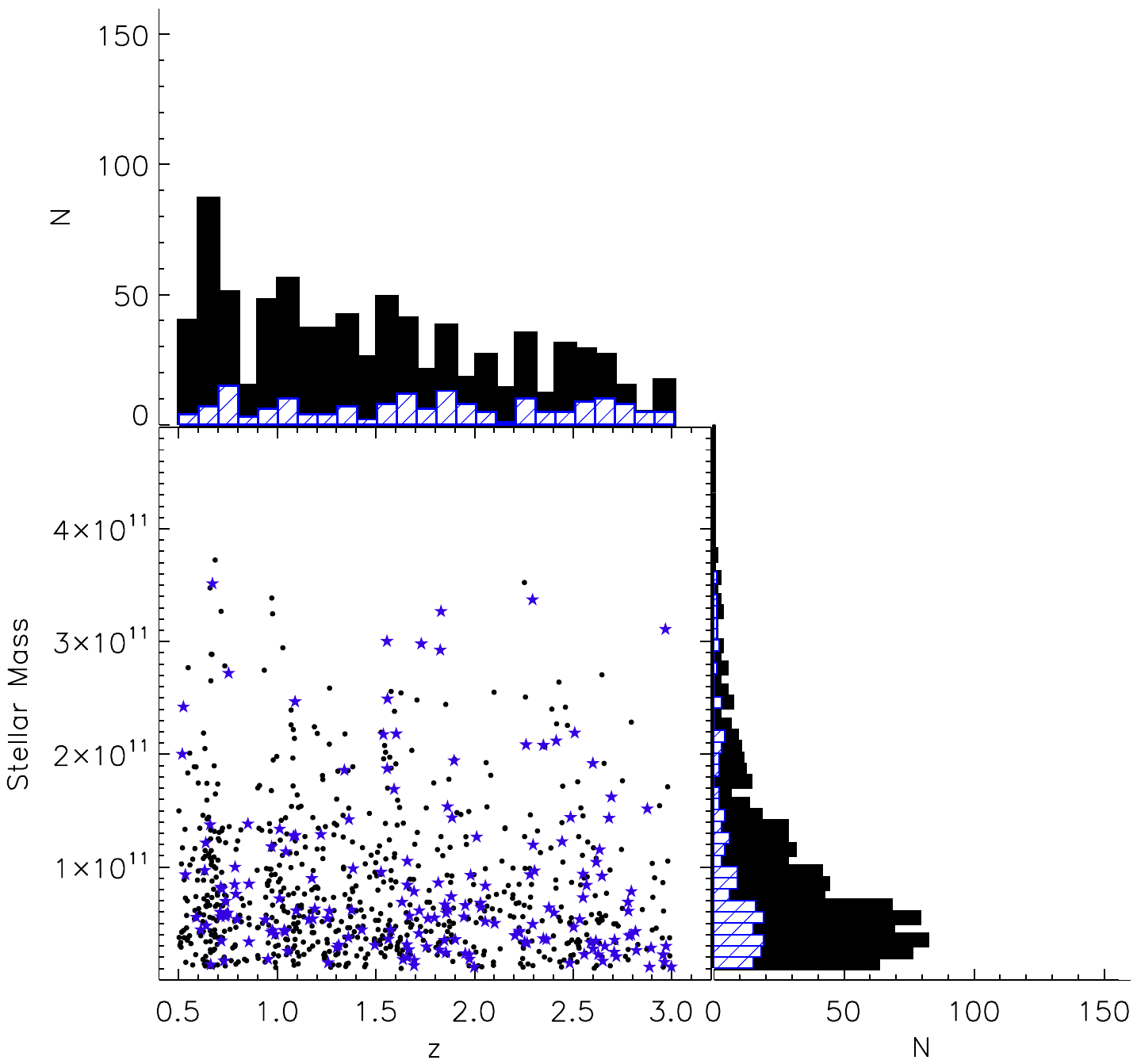}
 \caption{Mass and redshift distributions for the AGN host and mass-matched control samples. In the central panel we show the two dimensional distributions of the masses and redshifts for the X-ray selected AGN host sample in blue and the maximally selected non-active, mass-matched control sample in black. In the top histogram we display the redshift distributions for the control sample in black and the AGN hosts in blue, and likewise in the right histogram we plot the stellar mass distributions for the two samples.}
 \end{center}
 \end{figure}

In this way we ensure that the uncertainties in the comparison between the control sample and the AGN host sample are driven predominantly by the relatively low number of AGN hosts in each mass bin relative to the control sample. For all further analysis concerned with the comparison between AGN hosts and non-active galaxies, the distributions of physical properties of the control galaxy sample are given by the median of the 1000 bootstrapped samples within each bin for the different properties.

The mass and redshift distributions of our matched control sample and AGN hosts are shown in Fig. 1.
 
 Finally, we perform some additional cross-checks with AGN selected in different ways. First we compare to a sample selected based on the observed variability of their optical photometry by \citet{Falocco2015}. Given the relatively small area of the CANDELS GOODS-S field there is only a small overlap between the samples, but we find that those selected from optical variability within our mass and redshift range were already included in the \citet{Hsu2014} catalogue. We also compare to a sample of Compton-thick AGN selected by \citet{delMoro2015} and find another 3 AGN which are not in our original sample but are within our control sample and are fitted with pure disk morphologies. We conclude that by using an X-ray selected AGN sample we are selecting the largest fraction of radiatively luminous AGN and do not include the Compton-thick AGN in our analysis as our aim in this work is to explore any potential biases in the morphology fits of AGN hosts from a contribution from the AGN itself, and a Compton-thick sample should be the least affected by this. As a result, our conclusions apply to this subset of X-ray selected AGN.

 \section{Morphological Analysis}
 For the morphological analysis presented in this paper, we implement the procedure developed by Bruce et al. (2012, 2014) using the two-dimensional light profile fitting code {\sc GALFIT} \citep{Peng2002,Penggalfit2010}. In brief, we conduct the morphological analysis in the WFC3 F160W $H_{160}$-band, the reddest band available in the CANDELS survey which best represents the majority of the assembled stellar mass in the systems we are probing. The PSF we adopt is a hybrid composed of a TinyTim model \citep{Krist1995} in the central regions with an empirical fit used in the outskirts (for more details of the hybrid PSF generation see \citet{vanderWel2012}). We conduct two different classes of morphological fits: the first includes single S\'{e}rsic and single S\'{e}rsic + point-source components; and the second involves a full decomposition into the separate bulge, disk and point-source components. Throughout this analysis we limit our fits to bulges with S\'{e}rsic indices locked at n=4 (de Vaucouleurs bulges) and disks with S\'{e}rsic indices fixed to n=1 (exponential profiles). We also fix the centroid of each component, and thus only allow the magnitude, effective radius, axial ratio and position angle of each component to be free parameters in the fitting. Background determinations are conducted externally from within fixed, blank apertures and are applied prior to the fitting with {\sc GALFIT}. In order to decompose our objects appropriately, by default we adopt the simplest best-fit model and only accept a more complex model where it is statistically motivated given a likelihood ratio analysis, similar to the BIC test \citep{Schwartz1978} adopted by other authors. In order to combat cases where the {\sc GALFIT} $\chi^{2}$ minimisation routine iterates towards local rather than global minima, we start each fit with multiple different initial conditions to better explore the $\chi^{2}$ parameter space for each fit.
 
 For the purposes of this paper we concentrate on only the $H_{160}$-band morphological fits and defer a full description of the multi-wavelength decompositions which allow us to conduct individual component SED fitting to analyse the physical nature of the point-source components to the companion paper (Bruce et al., in preparation), where we will discuss this in detail.

 From our previous analysis and work with recovering simulated objects from within the CANDELS images \citep{Bruce2014}, we report uncertainties on bulge and disk magnitudes of $\sim10\%$, and size measurement of $10-20\%$ for disks and bulges, respectively. We have also explored the recoverability of point source affected objects and the prevalence of ``false positive" adoptions of these centrally concentrated compact models. We find that our procedure does not preferentially favour the adoption of a point-source component to deal with complex bulge+disk systems. Less than $0.5\%$ of bulge+disk fits of varying $B/T$ light ratios and sizes ratios adopt a spurious point-source component when such a component is not included in the simulated object, and degeneracies between the point-source component and the bulge only become considerable when the simulated bulges are unresolved (with effective radii $<1$ pixel). Thus, we conclude that the point-source components fitted in this analysis are robust and genuine morphological features.

 \section{Morphological Comparison}
 
 As discussed in the introduction of this paper, detailed morphological AGN host galaxy studies are implicitly difficult due to the need for high resolution, deep, imaging. Within the redshift range probed by this study, the CANDELS survey \citep{Grogin2011, Koekemoer2011}  has provided the ideal dataset with which to conduct these types of studies. There are already several studies which address the morphological structure of moderate luminosity X-ray selected AGN with this dataset  and as they provide an ideal basis against which to compare our newly decomposed morphologies, they are discussed briefly here for completeness. We note that the study of \citet{Schawinski2011} was the first work to utilise the WFC3 imaging on { \it HST} for AGN host galaxy analysis, but due to their smaller field coverage we do not directly compare to those results here.
 
The first of the studies we compare to is \citet{Kocevski2012} which is concerned with the visual classification of the near-infrared morphologies of the host galaxies of the \citet{Xue2011} 4Ms Chandra X-ray selected AGN in the ERS and deep regions of the CANDELS GOODS-S field. The aim of that work was to explore whether, compared to a mass-matched control sample, the AGN hosts display more disturbed morphologies indicative of mergers, thought to be one of the main drivers of AGN activity within this period of cosmic time. The second paper with which it is instructive to compare our results directly is \citet{Rosario2015}, which again utilised the 4Ms Chandra X-ray catalogue of \citet{Xue2011} in the GOODS-S field and the 2Ms Chandra catalogue of \citet{Alexander2003} in the GOODS-N to study the morphologies of low and moderate luminosity X-ray selected AGN hosts and a mass-matched control sample. This work uses both visual and parametric morphology fits using single component light profiles, and in agreement with \citet{Kocevski2012} finds that whilst most AGN hosts exhibit a disk-like morphology, compared to the mass-matched control sample the AGN hosts have a higher bulge fractions. However, \citet{Rosario2015} note the presence of a centrally concentrated light excess which may drive the single component light profile fits to higher, and thus more bulge-like, S\'{e}rsic indices. This excess is red in colour and may be the result of the AGN itself contributing to the rest-frame optical photometry of the object. Thus, the higher bulge fractions fitted to the AGN hosts may be biased by a contribution from the AGN and as a result, the trend for AGN host morphologies to have a higher contribution from a bulge component than the mass-matched control sample may be an artefact of inadequate fitting. This issue is well known and several other studies have addressed this by adopting an additional point-source light profile component in their morphology fitting procedure (e.g. \citealt{Schawinski2011}). However, it has also been noted that such an approach can result in the over-estimation of flux to the point sources and so drive the main stellar component profiles towards spuriously disk-dominated fits.

The extent to which the higher bulge fractions in the morphologies of AGN hosts are biased by a nuclear contribution from the AGN, and how the implementation of a point source in their light profile fitting affects their fits represents one of the main questions addressed by this paper, in addition to how the morphologies of AGN hosts differ from a mass-matched control sample when decomposed into their individual bulge and disk components.

 \subsection{Single S\'{e}rsic Models}
 
 In order to provide a comprehensive study of the effects of the exact morphological fitting technique employed on the trends observed between AGN host galaxy morphologies and the non-active control sample, we start by presenting our results for the simplest case of a single component S\'{e}rsic light profile component fit and then introduce an additional point-source component to explore the first additional level of complexity.
 The results of our single S\'{e}rsic fits (blue) are shown in the top panels of Fig. 2 and are over-plotted onto the fits for our mass-matched control sample (grey) in each redshift bin to allow for direct comparison. In agreement with \citet{Rosario2015} for the same CANDELS GOODS-S dataset, we observe that at the highest redshifts probed here the AGN hosts display a statistically distinguishable S\'{e}rsic index distribution, where the AGN host distribution is centred on higher index values, meaning that the hosts have morphologies with larger bulge contributions than the non-active control sample.
  
 \begin{table*}
 \centering
  \begin{tabular}{m{2.5cm} m{2.5cm} m{2.5cm} m{2.5cm} m{2.5cm} m{2.5cm}}
  Parameter&
  $0.5<z<1$&
  $1<z<1.5$&
  $1.5<z<2$&
  $2<z<2.5$&
  $2.5<z<3$\\
  \hline
  n&
  0.075&
  0.477&
  $<$0.001&
  0.004&
  0.001\\
  ${\rm r_{e}}$&
  0.151&
  0.851&
  0.851&
  0.781&
  0.126\\
  axial ratio&
  0.106&
  0.461&
  0.004&
  0.09&
  0.175\\ 
 \end{tabular}
 \caption{The one dimensional K-S test p values for the distributions of the: S\'{e}rsic indices, effective radii and axial ratios of the AGN hosts and the mass-matched control galaxies from the single S\'{e}rsic model fitting. }
 \end{table*} 
 
  The K-S p values for AGN host and non-active control sample S\'{e}rsic index distributions in each redshift bin are listed in Table 1, where it can be seen that above $z\sim1.5$ the AGN hosts and the non-active control sample are distinguishable at the $>95\%$ confidence level.

  \begin{figure}
 \centering
  \includegraphics[scale=0.5, trim=3cm 12cm 0cm 10cm] {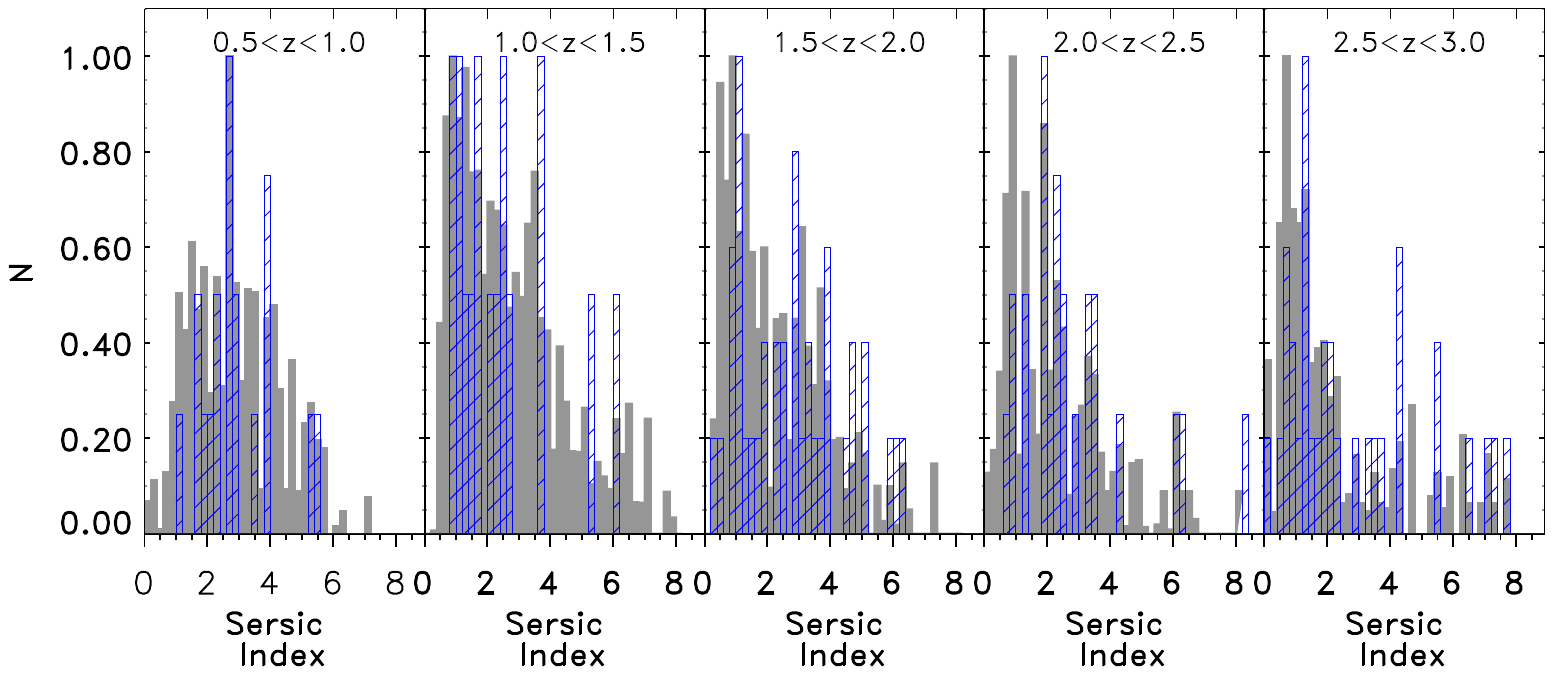}
  \includegraphics[scale=0.5, trim=3cm 12cm 0cm 10cm] {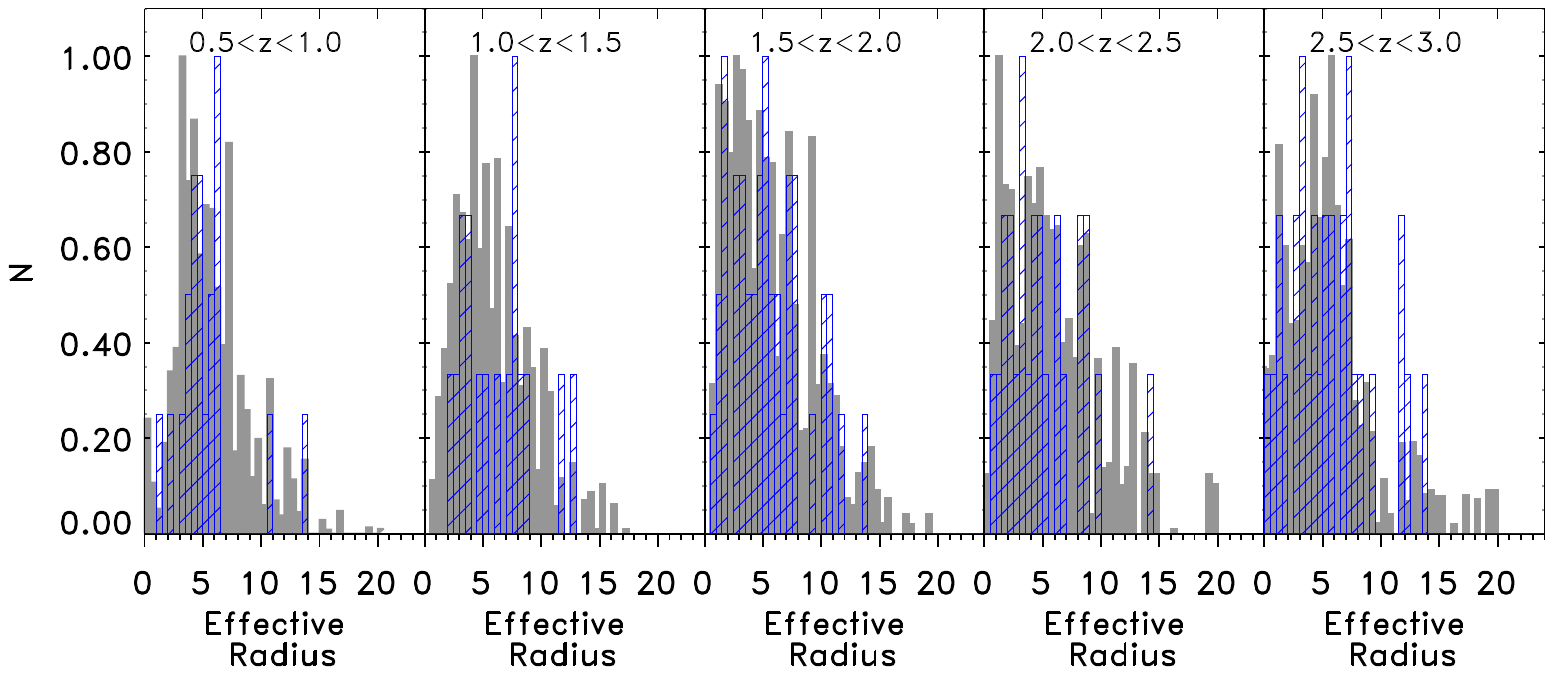}  
  \includegraphics[scale=0.5, trim=3cm 12cm 0cm 10cm] {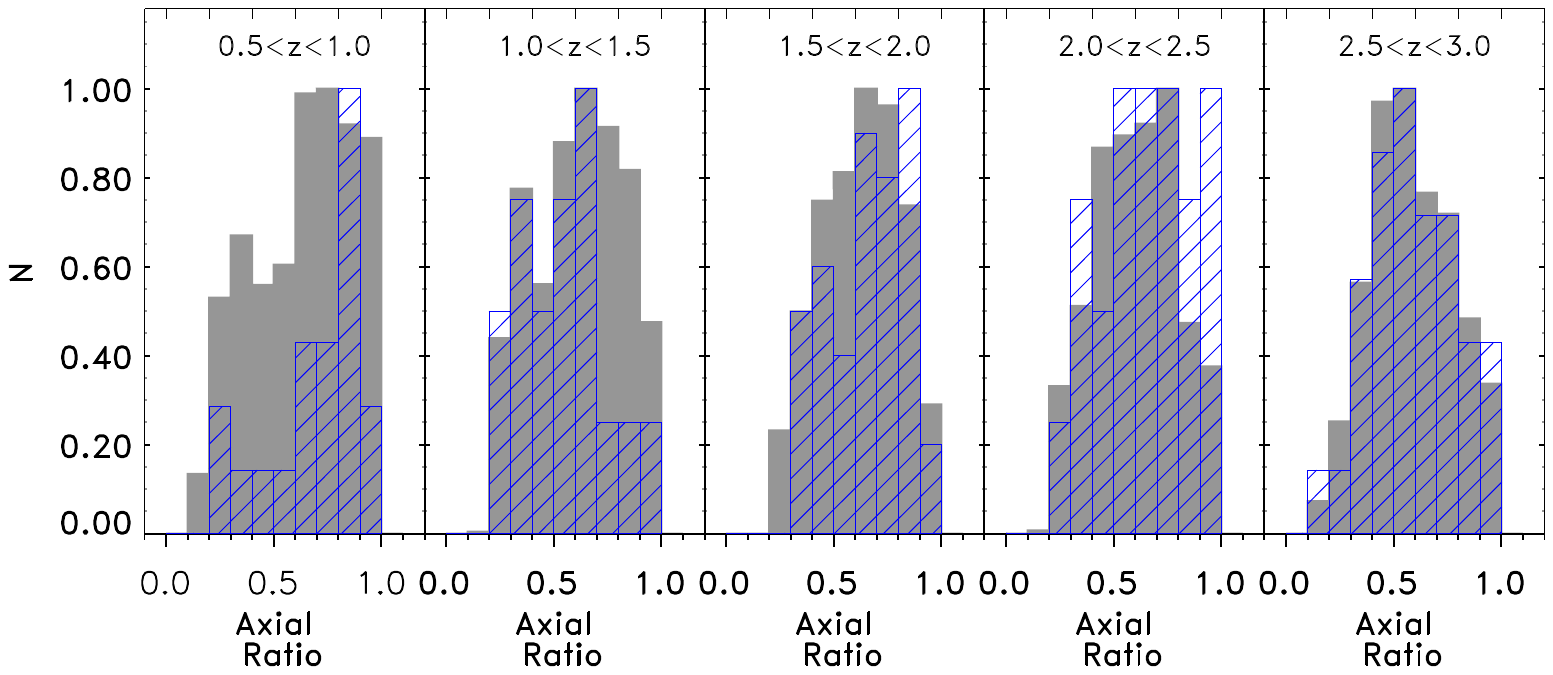}
\caption{The single S\'{e}rsic model fit parameter distributions for the control sample in grey, over-plotted by those for the AGN hosts in blue. The top row shows the S\'{e}rsic index distributions, the middle row shows the effective radii distributions and the bottom row shows the axial ratio distributions. These distributions have been normalised so that the most populated bin has a total of one. They reveal trends for the AGN hosts to have higher S\'{e}rsic indices at $z>1.5$. We do not find any significant evidence that the effective radii of the AGN hosts and control samples differ. However, we do report that the axial ratio distributions of the two populations are inconsistent in the $1.5<z<2$ bin, with the AGN hosts displaying a distribution with rounder systems than the control sample.}
 \end{figure}
 
 Within these panels the well-known evolution of massive galaxies within this redshift regime from disk to bulge-dominated \citep{Buitrago2008,Bruce2012} systems can also clearly be seen which, coupled with the comparison of the S\'{e}rsic index distributions of the AGN and control sample suggests that the AGN hosts may have an accelerated evolution, building up a significant bulge at earlier epochs compared to their mass-matched non-active analogues. However, overall it appears to suggest that the physical processes driving the evolution of these active and non-active systems are similar. Given the current evidence, from studies into the connection between galaxy star-formation rates and morphologies (e.g. \citealt{Bruce2012,McLure2013,Mancini2015}) in favour of a scenario where disks evolve through violent disk instabilities to gradually build up massive bulges rather than exclusively through gas-poor major mergers, this further suggests that major mergers may not play the dominant role in triggering AGN, as also commented on by \citet{Rosario2015}.
  
 A comparison of the effective radii and axial ratio distributions for the fits reveals no significant evidence for any offset in the effective radii distributions for the AGN host and non-active control samples. However, we do find evidence for a difference in the axial ratio distributions of the two samples in the $1.5<z<2$ redshift bin, where the AGN hosts display a distribution more dominated by rounder systems. This result agrees with our findings from the S\'{e}rsic index distributions that the AGN host sample has a larger bulge contribution than the control sample.

 \subsection{The Addition of Point-Source Models}
 
We next explore how the addition of a point-source component in our model affects the fitted morphologies of the AGN and the control sample. For this analysis we adopt the procedure implemented in \citet{Bruce2012}, where we only adopt a more complex model if it contains $\geq 10\%$ of the overall flux of the object and if the addition of the component reduces the $\chi^{2}$ value of the fit by more than expected given the increased degrees of freedom. Therefore, the results that we present in this section are updated only with the objects that require a point-source component in order to be better modelled. They do not illustrate the effect of imposing a point source in the models of all objects. In this way we allow the data to drive the level of morphological decomposition and do not make any {\it a priori} assumptions about certain samples, for example assuming that all AGN hosts should be fit with point-source components.

  From the best-fit models with the inclusion of a point-source component, we still see the same trend for AGN hosts to have higher bulge fractions than the control sample, where again the AGN host and control sample S\'{e}rsic index distributions become distinguishable at the $>95\%$ confidence level above $z=1.5$. The tabulated K-S p values for the AGN host and control sample S\'{e}rsic index, effective radii and axial ratio distributions, now for the best-fit single S\'{e}rsic + point-source fits, are given in Table 2, and the full distributions are plotted in Fig. 3. 
  
This can be seen in more detail when we look at the cumulative distributions of the two samples, as displayed in Fig. 4. From these plots it is clear that the differences between the S\'{e}rsic index distributions of the two samples occur due to the AGN hosts (in red) having relatively fewer low S\'{e}rsic index fits and more high S\'{e}rsic index fits compared to the non-active control sample.
Again, there is further evidence that the AGN hosts have a higher bulge fraction than the control sample from the comparison of the axial ratio distributions of the two populations in the $1.5<z<2$ redshift range, where the samples are inconsistent with being drawn from the same distribution at the $\sim 95\%$ confidence level.

As discussed above, by constructing a maximally mass-matched control sample and comparing the AGN host galaxy structural parameters to those of the medians of the 1000 times bootstrapped control samples, the statistical significance of these trends is measured with the relatively small number of AGN host galaxies within each redshift bin. As a result, the statistical significance of the trend for AGN hosts to have more bulge-dominated S\'{e}rsic indices than the mass-matched control sample, as measured by the p-value of the K-S test, can only be quoted at the $95\%$ confidence level, albeit in three consecutive redshift bins. The $\Delta z=0.5$ bins have been adopted for this study so as to ensure that there is no effect from morphological K-corrections within each redshift bin. However, by binning the full $1.5<z<3$ high-redshift sample together in order to increase the sample size of the AGN host galaxies, we find that we can reject the null hypothesis of the AGN host galaxy and control sample S\'{e}rsic indices being drawn from the same distribution at the $>99.7\%$ confidence level ($p<<0.001$).

  This confirms that the morphologies of the AGN hosts do indeed intrinsically have higher contributions from a bulge component than the control sample of non-active galaxies and this trend is not a result of any bias from overlooked centrally concentrated contributions from the AGN itself.
  
 \begin{table*}
 \centering
  \begin{tabular}{m{2.5cm} m{2.5cm} m{2.5cm} m{2.5cm} m{2.5cm} m{2.5cm} }
  Parameter&
  $0.5<z<1$&
  $1<z<1.5$&
  $1.5<z<2$&
  $2<z<2.5$&
  $2.5<z<3$\\
  \hline
  n&
  0.319&
  0.225&
  0.006&
  0.012&
  $<$0.001\\
  ${\rm r_{e}}$&
  0.341&
  0.901&
  0.197&
  0.836&
  0.085\\
  axial ratio&
  0.07&
  0.461&
  0.004&
  0.141&
  0.301\\ 
 \end{tabular}
  \caption{The one dimensional K-S test p values for the distributions of the: S\'{e}rsic indices, effective radii and axial ratios of the AGN hosts and the mass-matched control galaxies, now from the single S\'{e}rsic + point-source model fitting. }
 
 \end{table*}

 \begin{figure}
 \centering
  \includegraphics[scale=0.5, trim=3cm 12cm 0cm 10cm] {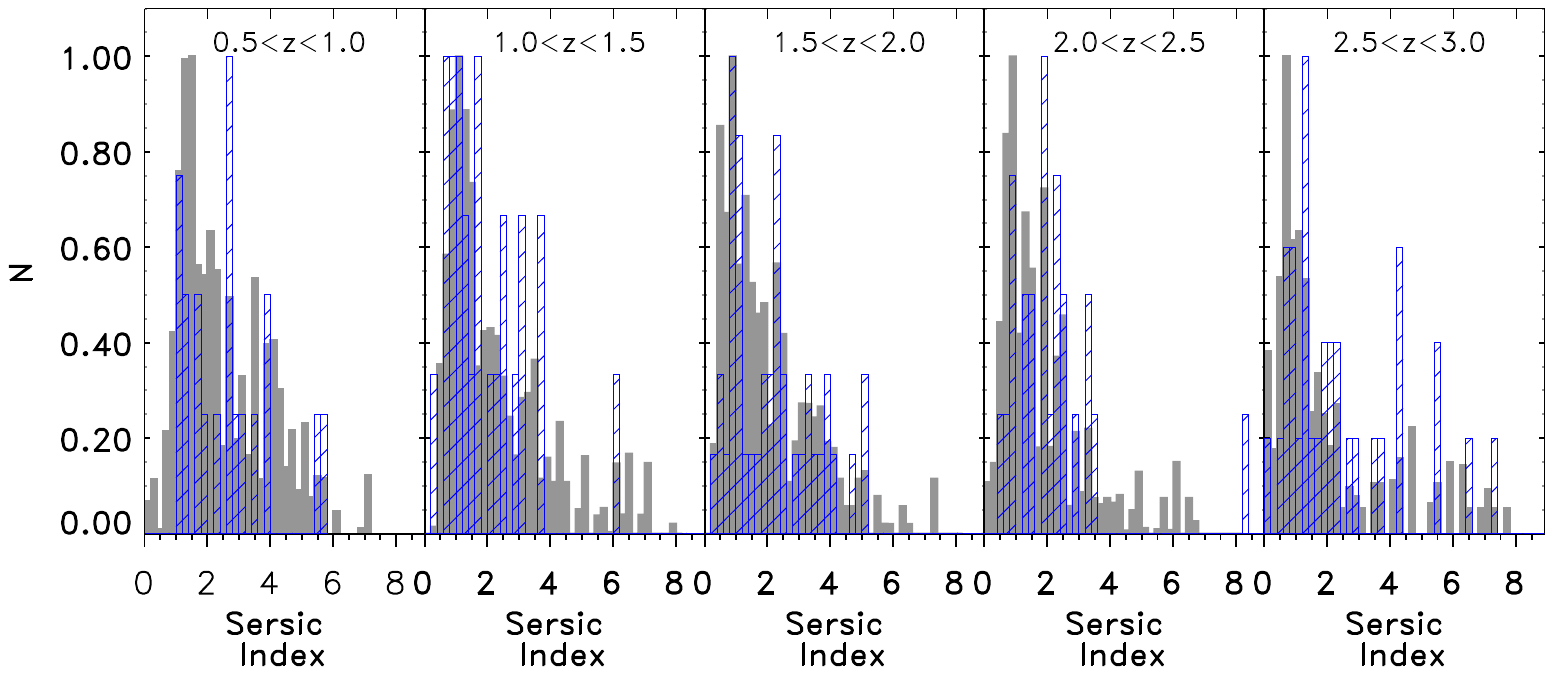}
  \includegraphics[scale=0.5, trim=3cm 12cm 0cm 10cm] {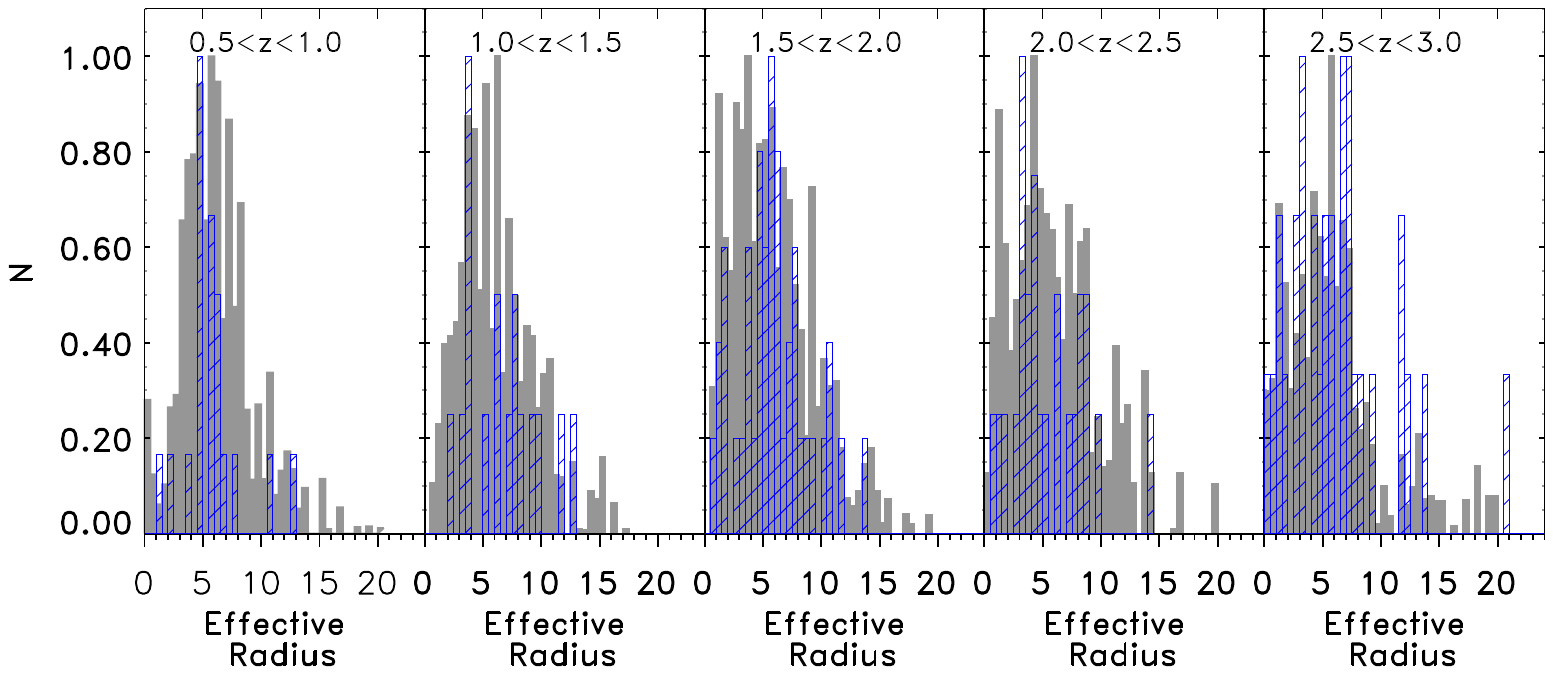}  
  \includegraphics[scale=0.5, trim=3cm 12cm 0cm 10cm] {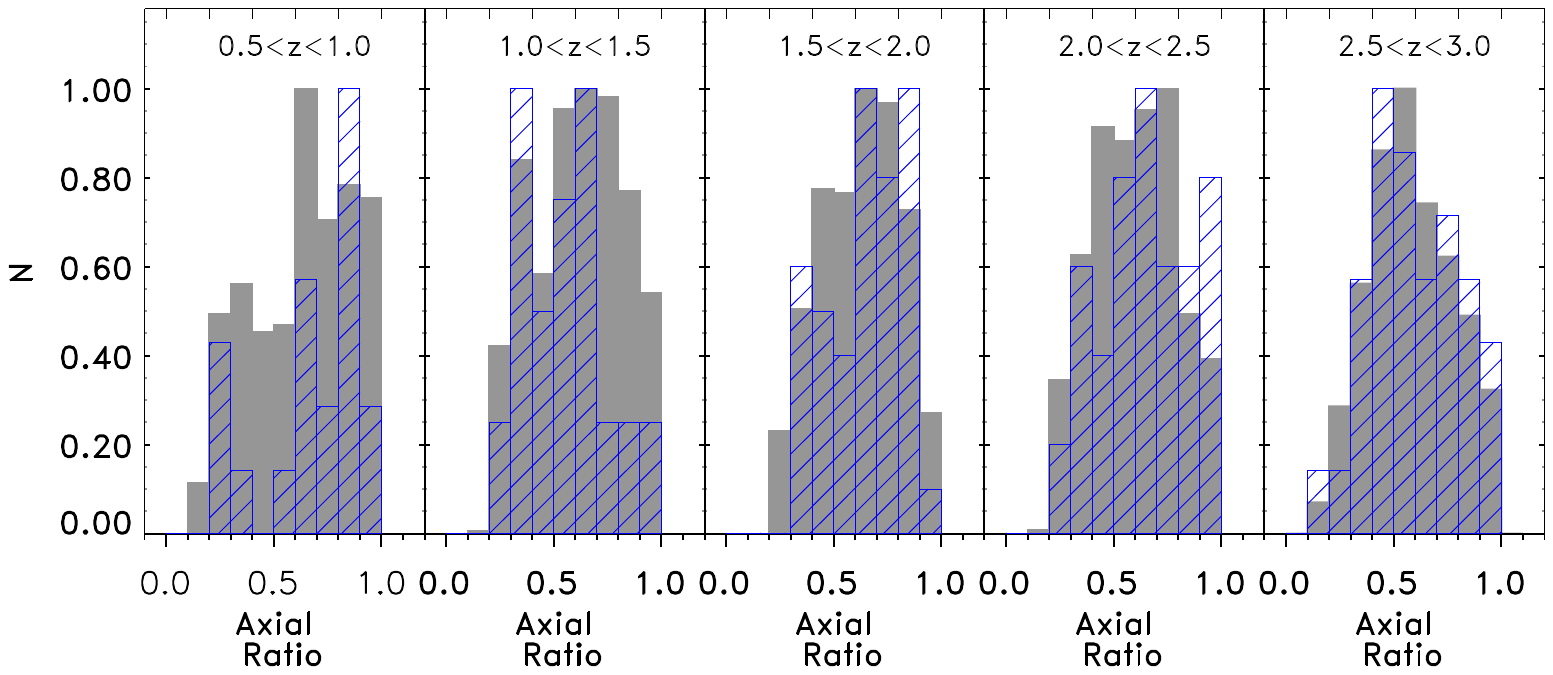}
\caption{In comparison to Fig. 2, these plots show the single S\'{e}rsic + point-source model fit parameter distributions. Again, the control sample results are plotted in grey and are over-plotted by those for the AGN hosts in blue. The top row shows the S\'{e}rsic index distributions, the middle row shows the effective radii distributions and the bottom row shows the axial ratio distributions. As with Fig. 2, these distributions have been normalised so that the most populated bin has a total of one. It is clear from a direct comparison with Fig. 2 that the same trends exists for the AGN hosts at $z\geq2$ regardless of the inclusion of the point-source component in the morphology fitting. However, the significance of these trends is weakened below $z=2$. }
 \end{figure}

\begin{figure*}
 \centering
  \includegraphics[scale=0.25, trim=0cm 13.cm 3cm 0cm] {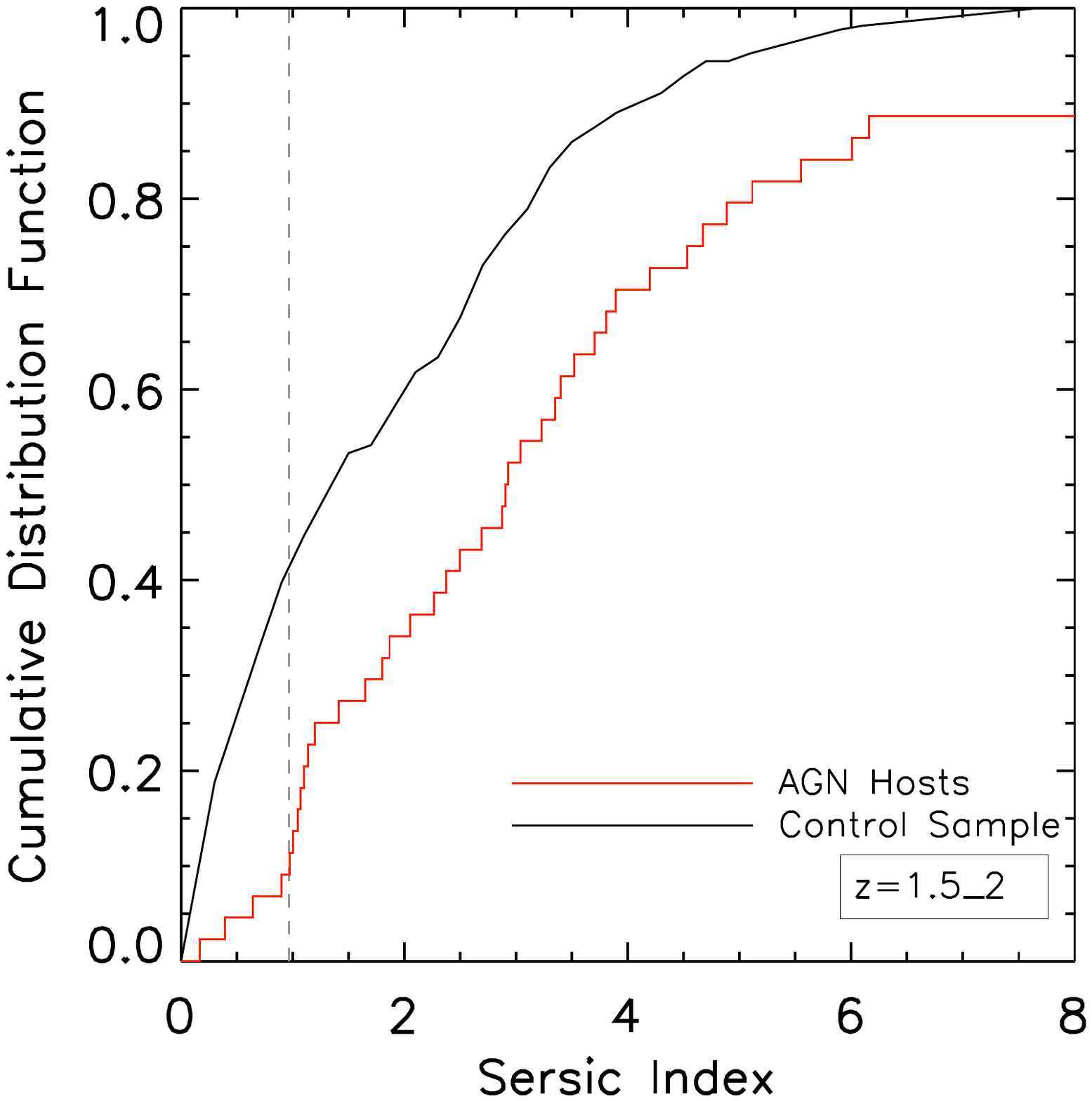}
 \includegraphics[scale=0.25, trim=0cm 13.cm 3cm 0cm] {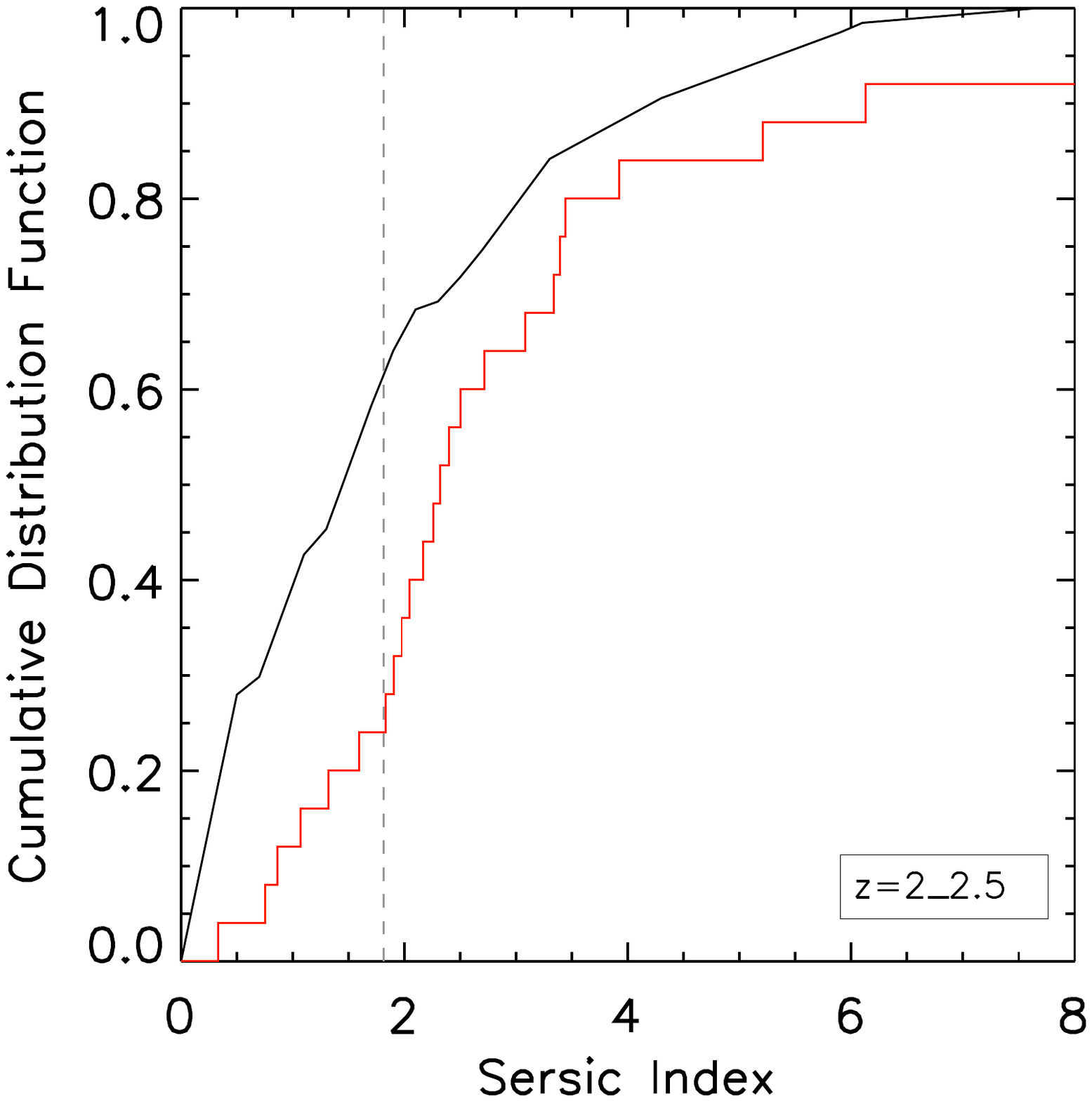}
 \includegraphics[scale=0.25, trim=0cm 13.cm 3cm 0cm] {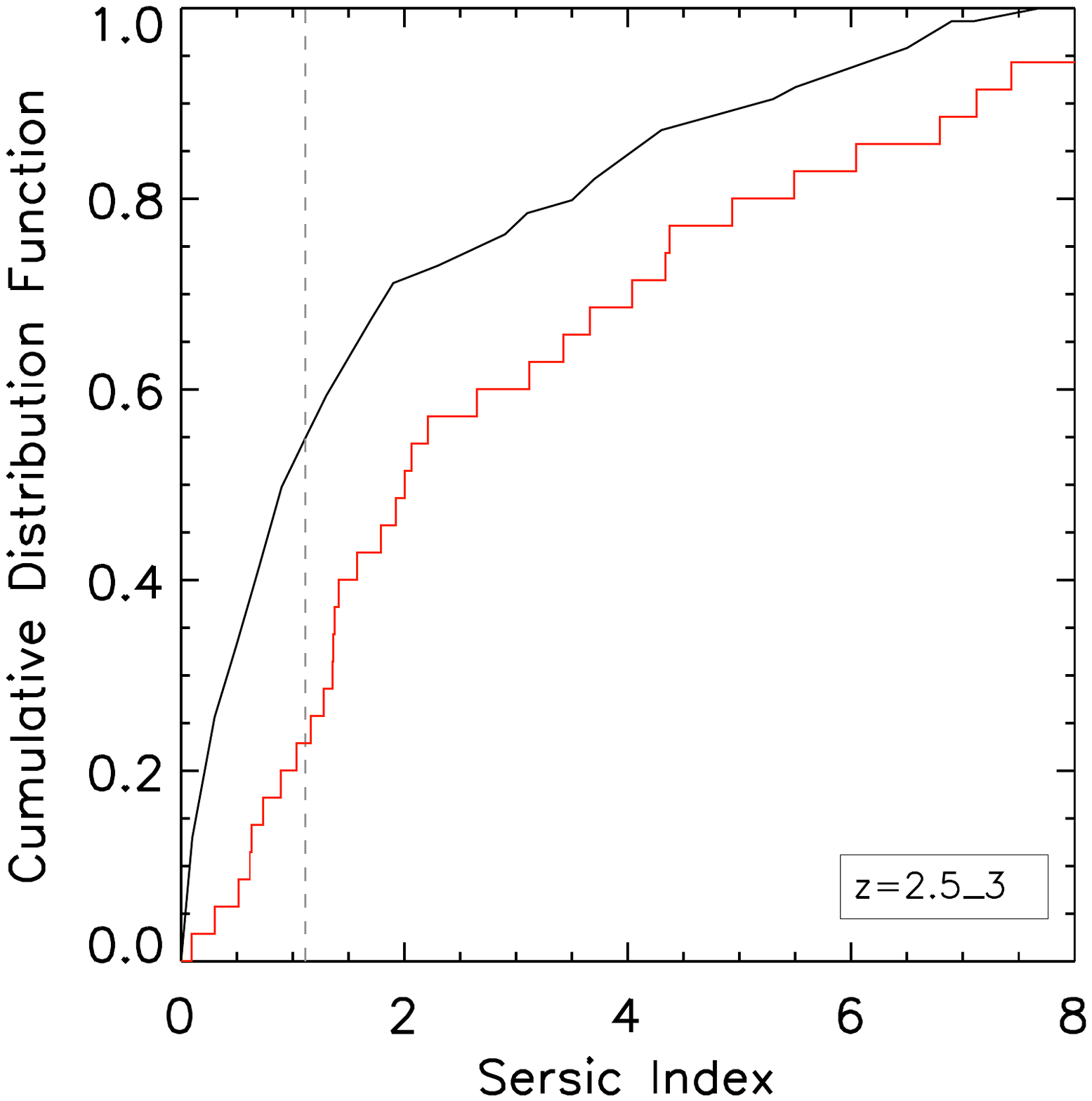}
 \includegraphics[scale=0.25, trim=0cm 13.cm 3cm 0cm] {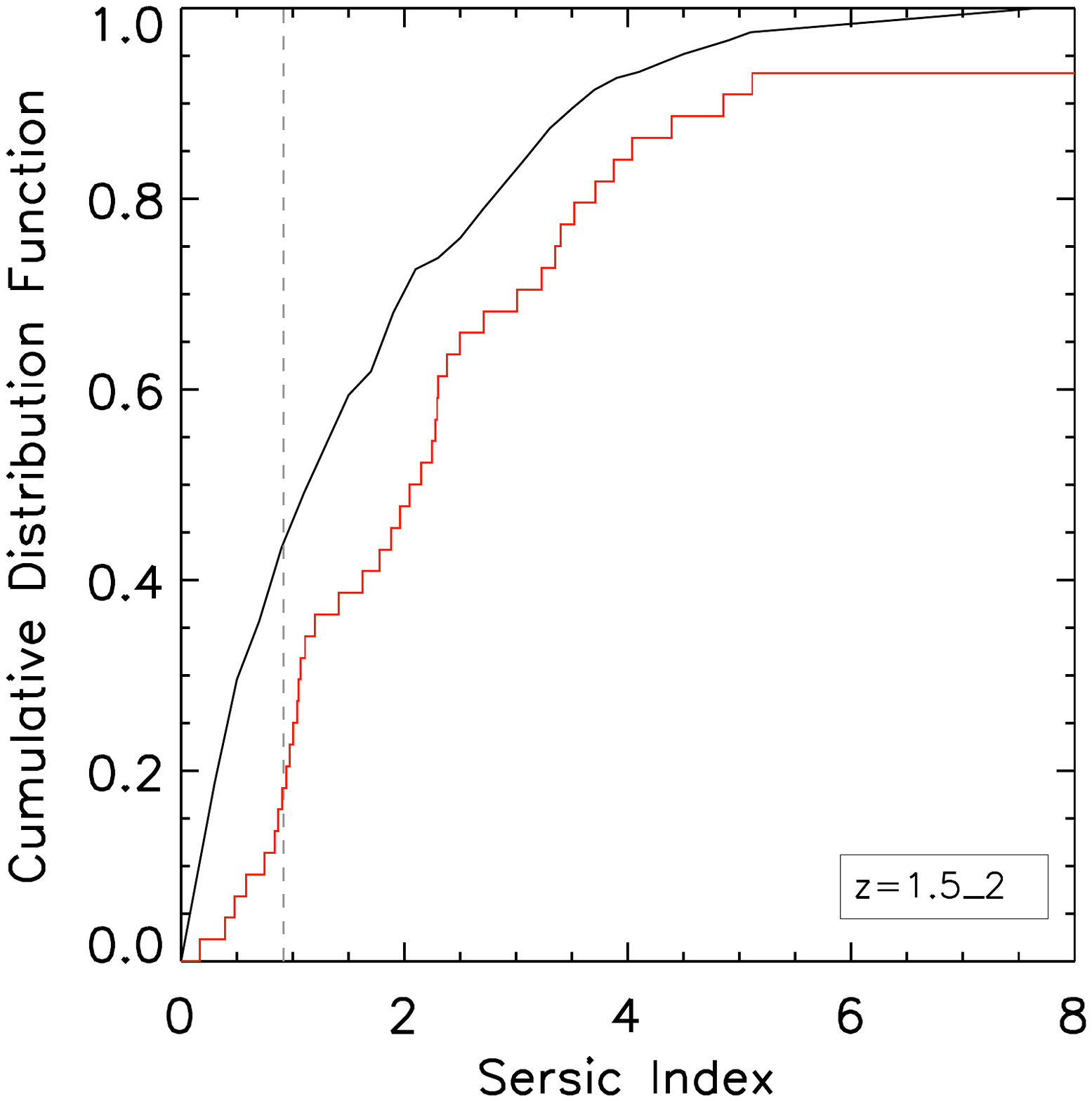}
 \includegraphics[scale=0.25, trim=0cm 13.cm 3cm 0cm] {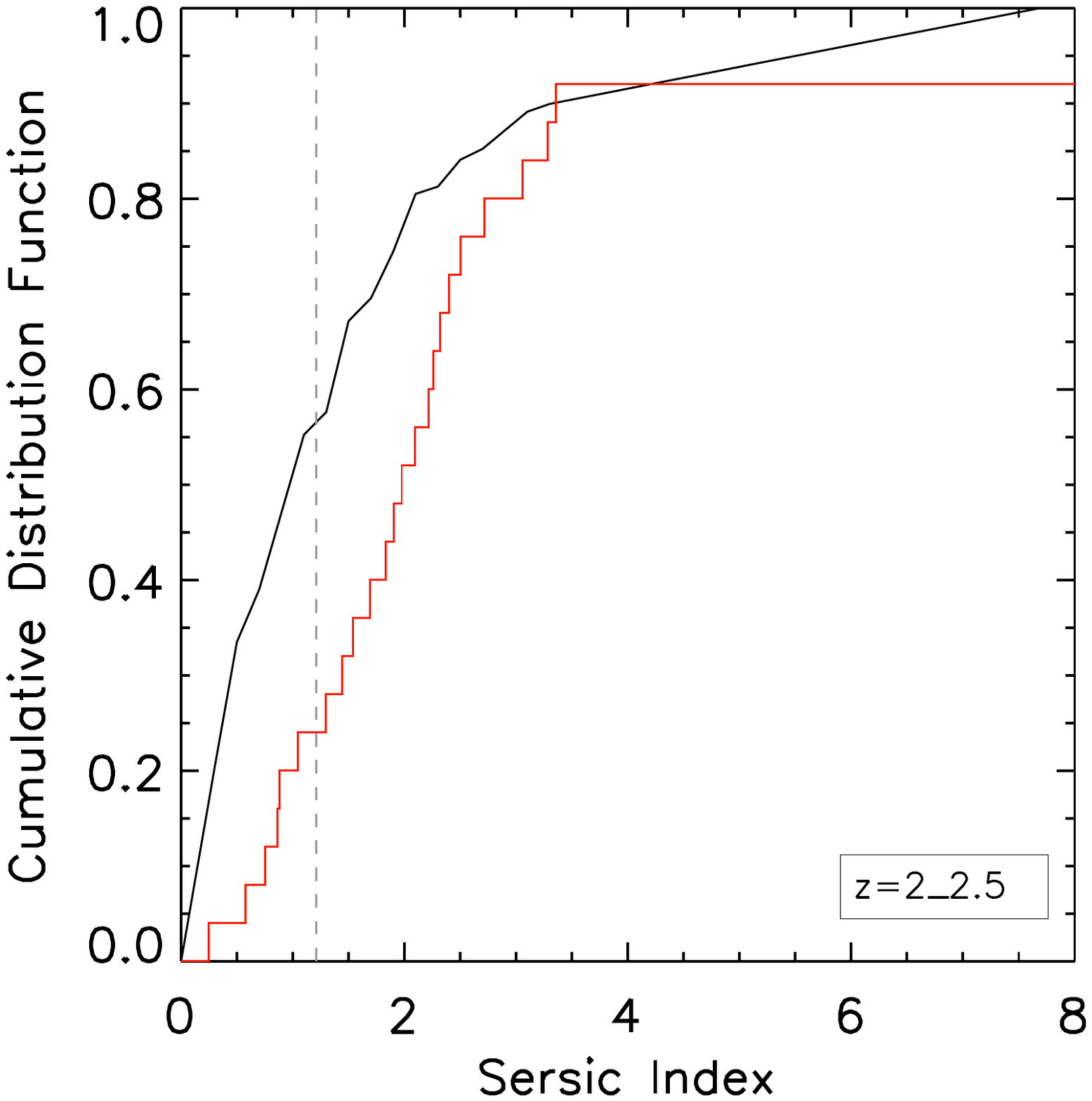}
  \includegraphics[scale=0.25, trim=0cm 13.cm 3cm 0cm] {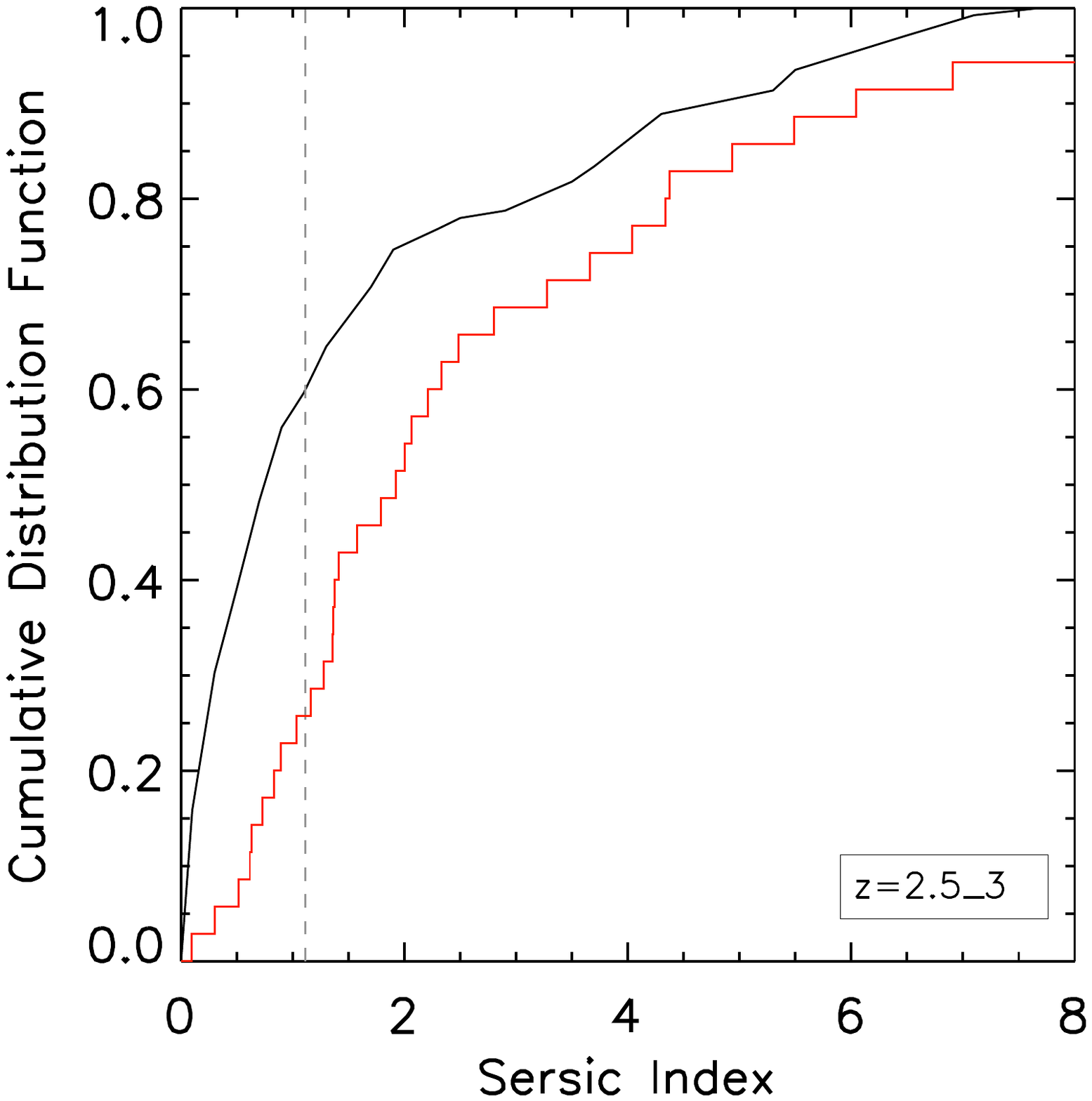}

  \caption{The cumulative distribution functions for the S\'{e}rsic indices for the single S\'{e}rsic best-fit models on the top row and for the best-fit single S\'{e}rsic +point-source models in the bottom row. The AGN hosts are plotted in red and the control sample is over-plotted in black. The indices at which the largest deviation between the distributions occurs are marked by the vertical dashed, grey line. We note that the maximum S\'{e}rsic index allowed during the fitting was n=20, but for the ease of this comparison the S\'{e}rsic indices have only been plotted out to n=8. These plots highlight that the largest differences in the distributions between the active and non-active samples occur at low S\'{e}rsic indices, where the AGN hosts display a lower overall contribution from low index (i.e. more disk-dominated) fits, indicating that the AGN host structures have a larger contribution from a bulge component.}
 
 \end{figure*}

 \subsection{Multiple S\'{e}rsic Models}
 
 Given the differences in the morphologies of the AGN hosts and the control sample indicated from a single S\'{e}rsic fit to the data, we now explore these trends in more detail with a full bulge-disk decomposition of both samples.
The methodology for choosing best-fit models from our decompositions was previously developed for a high mass (M $>10^{11}{\rm M_{\odot}}$) sample with higher $S/N$ than the M $>10^{10}{\rm M_{\odot}}$ sample selected for this work. As a result, when applied to this sample we found that the $\chi^{2}_{complex} <\chi^{2}_{simple}-\Delta\chi^{2}(\nu_{complex}-\nu_{simple})$ criteria imposed to adopt more complex models was too stringent and returned best-fits which were limited to single components. While these fits may be the most statistically validated fits on an object-by-object basis, it is clear they do not best represent the morphologies of the overall population. For this reason, for the multiple component fits for this lower mass sample of both AGN hosts and the non-active control sample we relax the $\chi^{2}$ criteria and simply adopt as best-fit the model with the lowest $\chi^{2}$ value, but still ensure that any component is $\geq 10\%$ of the overall flux of the object. Having tested this approach on the high-mass end of this sample, we have ensured that the relaxation of the $\chi^{2}$ criteria does not bias the fits towards any non-physical model set-up, for example disk-dominated bulge+disk objects are best-fit with pure disks when the criteria are strengthened and vice versa.

 \begin{figure}
 \centering
  \includegraphics[scale=0.5, trim=2.5cm 12cm 0cm 10cm] {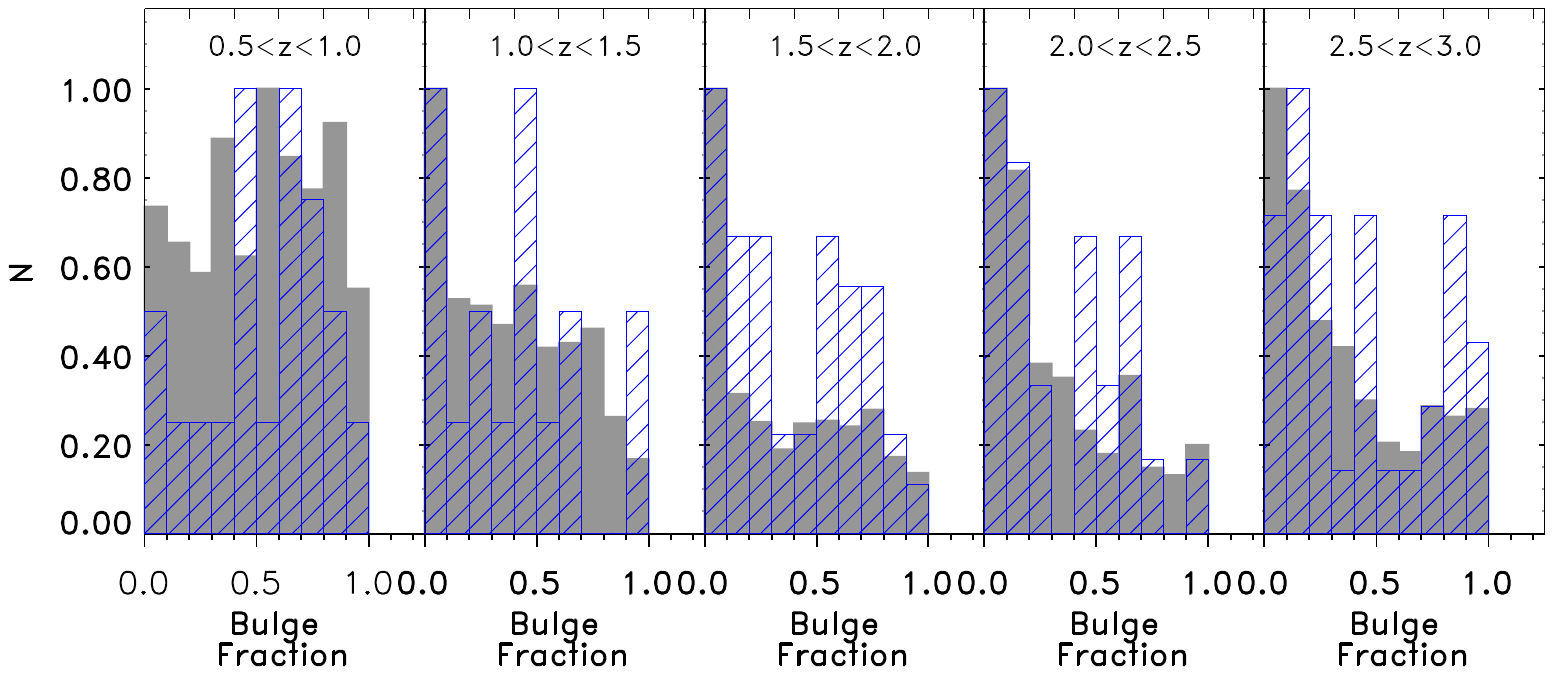}
  \includegraphics[scale=0.5, trim=2.5cm 12cm 0cm 10cm] {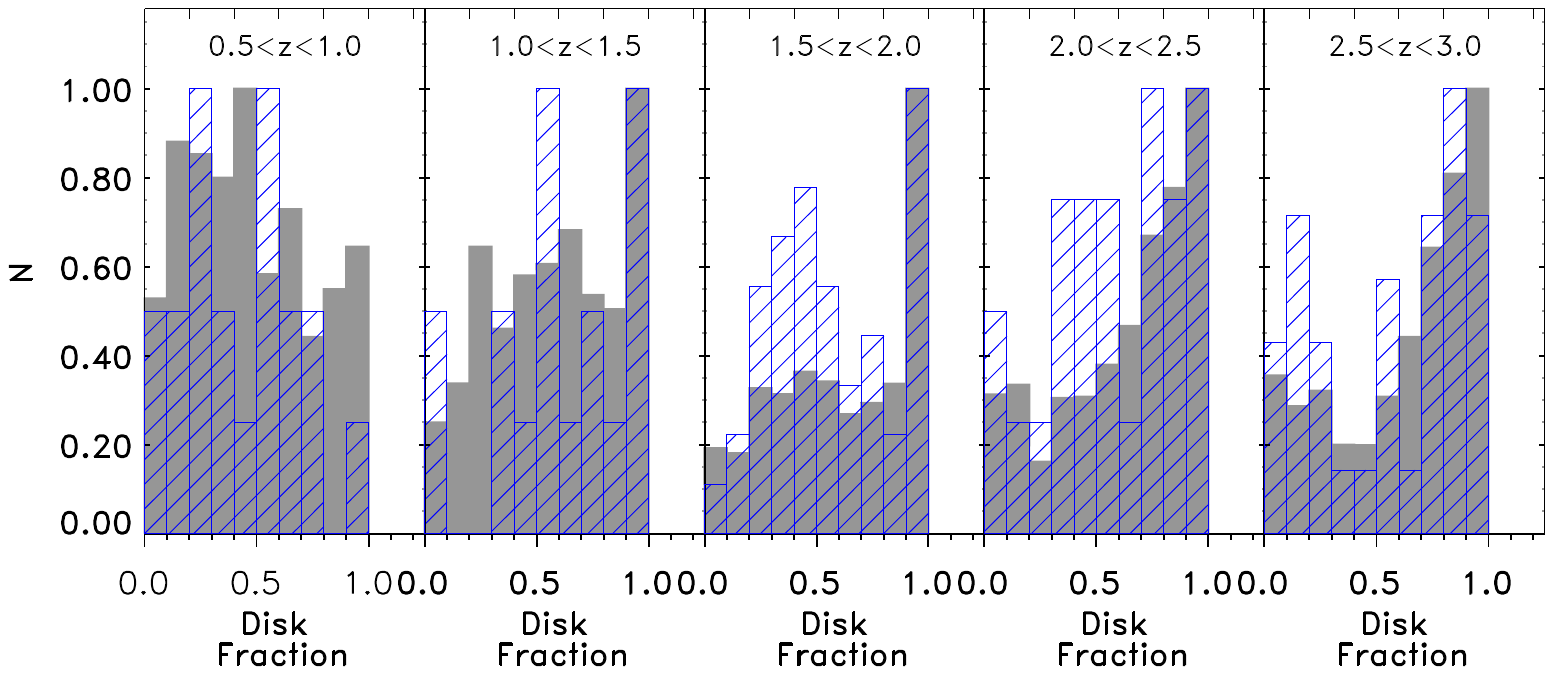} 
\caption{The light fraction distributions from our full morphological decomposition for the control sample in grey, with the distributions for the AGN hosts over-plotted in blue. The bulge component light fractions are given in the upper row and the disk component in the lower row. Above $z=2.5$ these distributions reveal that the AGN hosts have fewer low bulge components and more lower disk components than the control sample. However, these trends are more easily seen in the cumulative distribution functions shown in Figs. $6 \& 7$.}
 \end{figure}

The bulge and disk light fraction distributions obtained from this approach are shown in Fig. 5. Again, here the results presented depict the best-fit morphologies which have the option to adopt a point-source component where motivated by the data.

From a one-dimensional K-S test of the AGN hosts and the median of the 1000 bootstrapped mass-matched control samples, we note that the bulge light fractions become distinguishable at the $95\%$ confidence level above $z=2.5$. From the cumulative distribution functions, Fig. 6 for the bulge light fractions and Fig. 7 for the disk light fractions, we can see that the AGN host sample displays fewer low bulge fraction components and more low disk fraction components, in additional to more pure bulges systems. Therefore, this further corroborates our findings that, even when fully decomposed, the AGN hosts have higher bulge fractions than a mass-matched non-active control sample.

 \begin{figure}
 \centering
  \includegraphics[scale=0.35, trim=0cm 13cm 0cm 0cm] {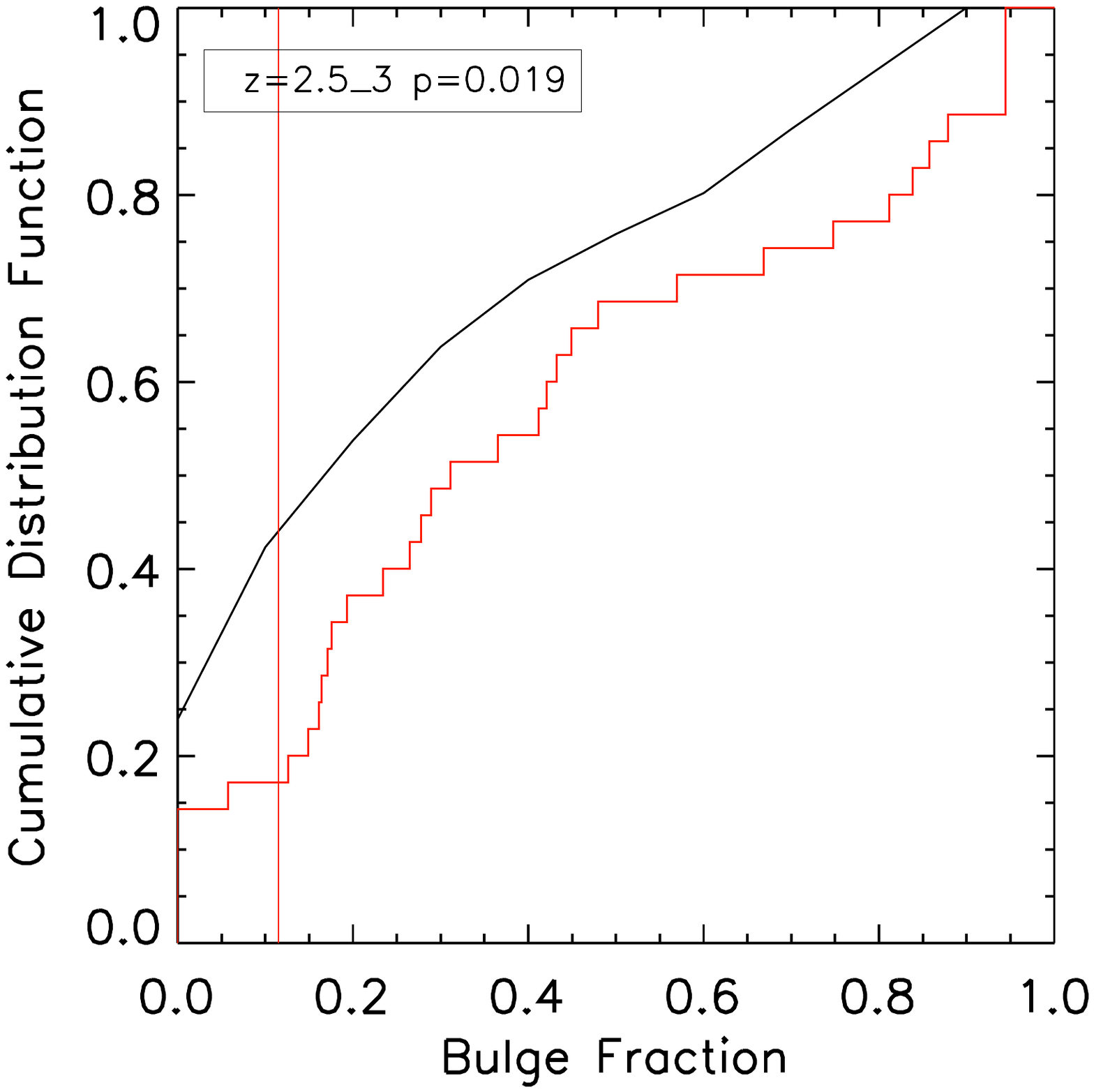}
\caption{The cumulative distribution function for the bulge component light fractions. The AGN hosts are given in red, and the control sample light fractions are shown in black. The red vertical line represents the bulge fraction at which the biggest difference between the cumulative distribution functions occurs. From these plots it is clear that the differences in these distributions arises due to the AGN hosts having fewer low bulge fraction components, and also more high bulge fraction components.}
 \end{figure}
 \begin{figure}
 \centering
  \includegraphics[scale=0.35, trim=0cm 13cm 0cm 0cm] {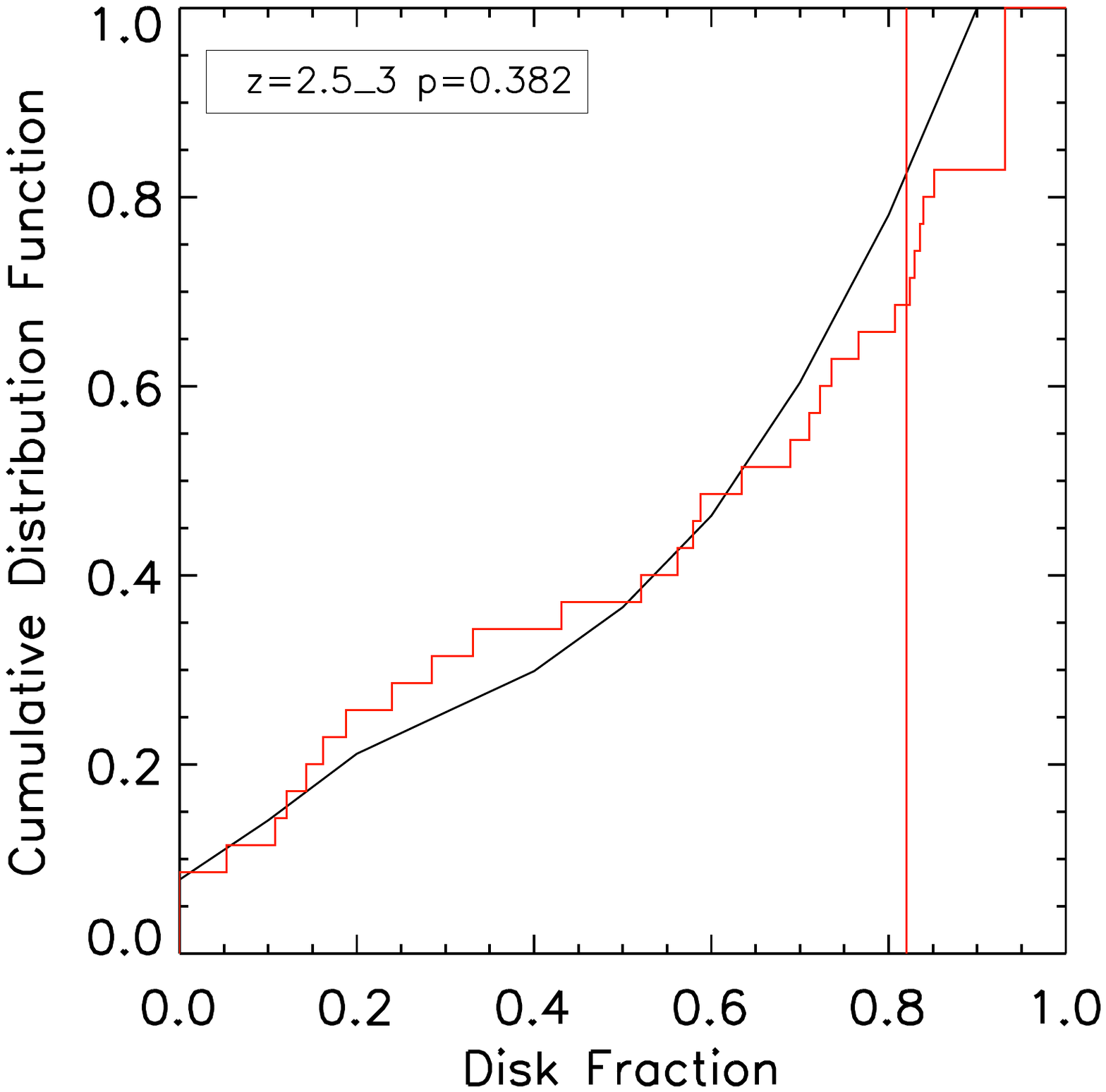}
\caption{Following on from Fig. 6, this plot shows the cumulative distribution function of the disk component light fractions. Again the distribution function of the AGN hosts is plotted in red and that of the control is in black. The red vertical line represents the disk fraction at which the biggest difference between the cumulative distribution functions occurs. Comparison of these distributions reveals that in addition to the AGN hosts having more low disk fraction components, they also have fewer high disk fraction components, consistent with the claim that the AGN hosts have higher bulge fractions than the non-active control sample.}
 \end{figure}

Finally, we address the issue of morphological K-corrections within our $0.5<z<3$ range. It is clear that the adoption of our $H_{160}$-band morphological classifications probes increasingly bluer rest-frame light at higher redshifts. It is possible that this may contribute to the higher bulge fractions observed in the AGN host sample compared to the control sample if the contribution to nuclear light from the AGN itself becomes more dominant at the bluer wavelengths probed by our fixed-band morphological decomposition at higher redshifts. However, from our detailed multi-wavelength decompositions we do not find any such differences in the broad-band SEDs of the control sample galaxies compared to the X-ray identified AGN hosts for the majority of objects. Thus, we conclude that there is no evidence to suggest that the $H_{160}$-band morphologies of the AGN hosts are more biased by contributions from the AGN itself at higher redshifts, and so we do not consider this as a major concern.
Moreover, we find no trend for the point-source fits to be adopted with higher frequency, or to have higher point-source fractions, at high-redshifts compared to lower redshifts.
 
   \subsection{The Role of the Galaxy Bulge in Determining AGN Activity}
 Given that we have determined robust $H_{160}$-band light fraction decomposed stellar masses for our AGN sample, we are also able to address the issue of whether the BH mass shows a stronger trend with the bulge as opposed to the total stellar mass of the host.
 Determining BH mass estimates from dynamical measurements is beyond the scope of this work. However, we can further exploit our comparison between the total masses and decomposed bulge masses of our AGN hosts and of our full $M_*>10^{10}{\rm M_{\odot}}$ sample in order to shed some light on this issue.
 
  \begin{figure}
 \centering
 \includegraphics[scale=0.55, trim=0cm 11cm 0cm 6.8cm] {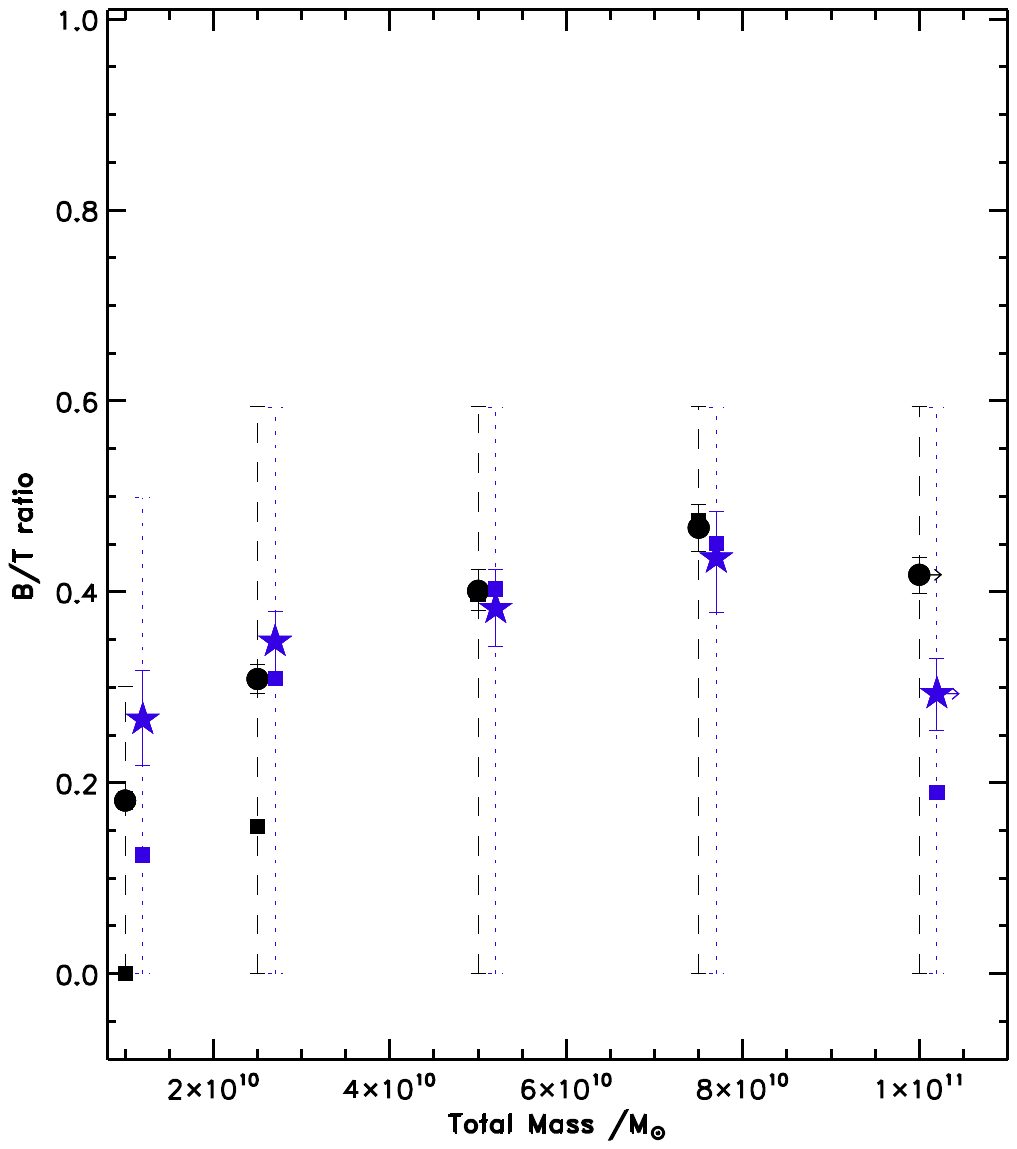}
\caption{The binned bulge fractions of the AGN and the full CANDELS GOODS-S, $0.5<z<3$, $M_*\geq1\times10^{10}{\rm M_{\odot}}$ non-active galaxies as a function of the total stellar mass in the systems. Here the mass bins are $\Delta M_*=2.5\times10^{10}{\rm M_{\odot}}$ in size and extend to $M_*=1\times10^{11}{\rm M_{\odot}}$ at which point all higher mass systems are included in the $M_*\geq1\times10^{11}{\rm M_{\odot}}$ bin in order to maintain a reasonable number of objects in each bin. The AGN hosts are displayed in blue and the full non-active, $M_*>10^{10}{\rm M_{\odot}}$, sample is shown in black. The large circles represent the mean bulge fractions in the bin, whereas the small squares are the median values. The interquartile ranges are also given by the dotted and dashed lines for the AGN and non-active galaxies, respectively, and have been offset to allow for clearer viewing. From this comparison we can see that the lower mass AGN hosts have a higher mean and median bulge fraction than the non-active control sample. }
 \end{figure}
 
 In Fig. 8 we present the binned bulge fractions of the AGN hosts (in blue) and the full sample (in black) as a function of the total stellar masses of the systems. Both the mean (large circles), median (small squares) and interquartile ranges have been displayed in the figure to allow a full comparison of the distribution of bulge fractions for the two populations. As a reminder, we have selected this sample via a mass and photo-z selection cut at $ M_*>1\times10^{10}{\rm M_{\odot}}$ with $0.5<z<3$ , and have cross-matched these objects with the full \citet{Hsu2014} 4Ms Chandra X-ray selected counterpart catalogue. Thus, the higher bulge fractions displayed in the lower mass AGN hosts compared to the non-active galaxies suggest that in order to have been included in an X-ray flux limited catalogue, the lower mass AGN hosts had to have a larger contribution from a bulge component. When viewed in light of the strong trend between bulge mass and total stellar mass, this shallower evolution in bulge mass with total stellar mass for the low mass AGN hosts compared to the non-active sample indicates that the X-ray luminosity, and therefore BH mass, is more correlated with the bulge stellar mass of the hosts than the total mass.
  
   \begin{figure}
 \centering
 \includegraphics[scale=0.44, trim=1.5cm 12cm 0cm 7cm] {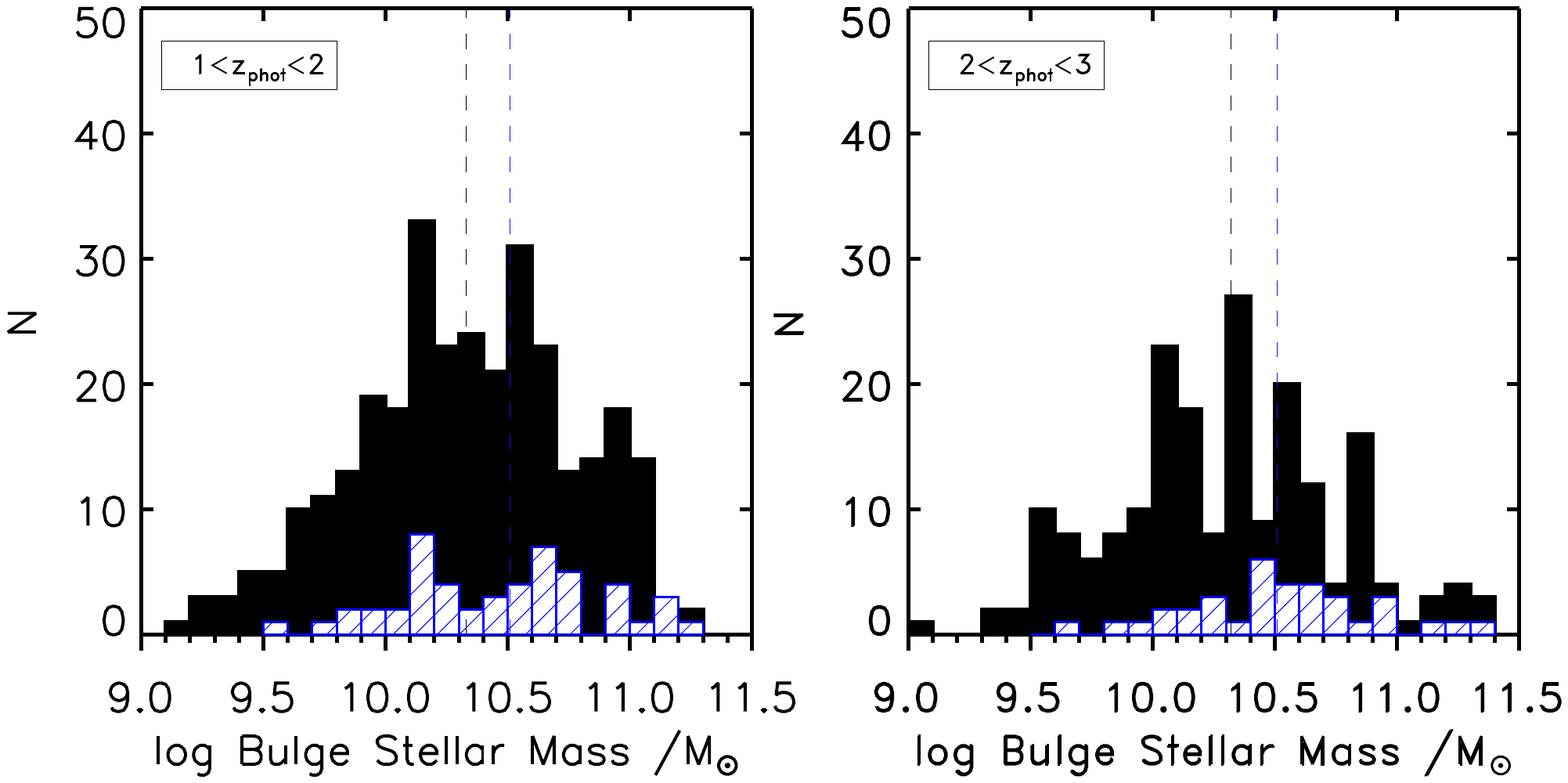}
\caption{ The bulge stellar mass distributions for the full CANDELS-GOODSS, $0.5<z<3$, $M_*\geq1\times10^{10}{\rm M_{\odot}}$ galaxy population in black, over-plotted with the AGN host sample in blue, at two different redshift intervals. The medians of the two distributions have been over-plotted in the blue and black dashed lines for the AGN hosts and control sample respectively. These distributions indicate that the AGN hosts have fewer low bulge stellar mass systems, thus suggesting that the $ M_*=1\times10^{10}{\rm M_{\odot}}$ mass-cut maximises the overlap with the X-ray selected AGN host sample of \citet{Hsu2014}  due to the fact that the inclusion of $ M_*>1\times10^{10}{\rm M_{\odot}}$ bulge stellar masses systems is needed to properly sample the AGN host galaxy population. Furthermore, this trend in bulge stellar masses may also play an important role in the observation that generally AGN hosts are biased towards higher total stellar masses within a more general mass-cut galaxy sample.}
 \end{figure}
 
 This exploration of the bulge and total stellar mass distributions for the AGN hosts and control sample also provides clarity on the effectiveness of our $ M_*>1\times10^{10}{\rm M_{\odot}}$ galaxy cut at encompassing the majority of the X-ray identified AGN from the counterpart catalogue of \citet{Hsu2014}. As we can see from Fig. 9, within the scatter, the bulge stellar mass distribution of the AGN host population truncates at higher bulge mass than the overall galaxy population. In addition to this, looking at the overall distributions, the AGN hosts have higher median bulge stellar masses ( $\log_{10} (M_*/{\rm M_{\odot}} )= 10.51$ in both redshift bins) compared to the control samples ($\log_{10}(M_*/{\rm M_{\odot}})=10.33$ and $\log_{10} (M^*/{\rm M_{\odot}} )= 10.32$ for the low and high-redshift bins respectively). Thus, it is clear that systems with bulge masses $>1\times10^{10}{\rm M_{\odot}}$ are needed in order to fully represent the AGN host population within a mass-matched sample and in fact this correlation with higher bulge masses may be the reason for the well-known trend for AGN hosts to be biased towards higher total stellar masses than the underlying galaxy population within a given mass-cut sample.

  \section{Nature of the Point-Source Component}
 
 Having determined that the trend for X-ray selected AGN host galaxy morphologies to be more bulge dominated at high redshifts is not an artefact of the poor fitting of any contribution from the AGN itself to the near-infrared photometry of these objects, we now turn our attention to the physical nature of the point-source component adopted in some of the fits, and ask how well this correlates with the presence of an AGN.

\subsection{Point-Source Fraction Differences}
 It is clear from this, and our previous \citep{Bruce2012,Bruce2014} studies, that a significant fraction of massive galaxies require a point-source component in order to best-fit their light profiles. So how does the fraction of point-source fits in the mass-matched control sample compare to that in the X-ray selected AGN hosts ?
 In Fig. 10 we show the cumulative distribution functions of the point-source light fractions in each redshift bin for the AGN hosts in red and the mass-matched control sample in black, and over-plot in each panel the one dimensional K-S test p values for the distributions. This demonstrates that, overall, we fail to reject the null hypothesis that the distributions of the point-source fractions are drawn from the same distribution, thus revealing that there is no evidence for the AGN hosts to be more point-source dominated than the non-active control sample.

\begin{figure*}
 \centering
  \includegraphics[scale=0.2, trim=0cm 13.5cm 5cm 0cm] {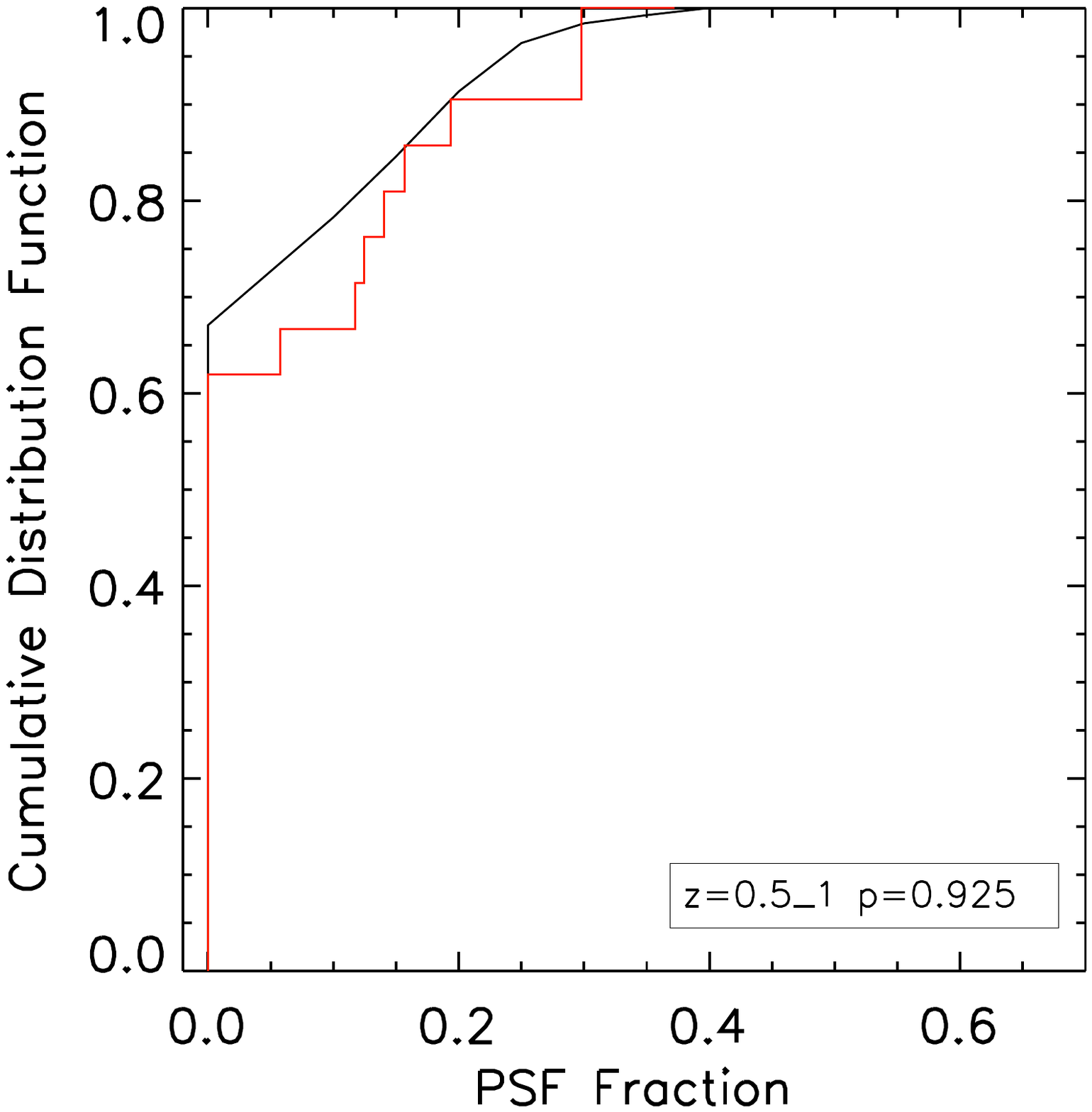}
 \includegraphics[scale=0.2, trim=0cm 13.5cm 5cm 0cm] {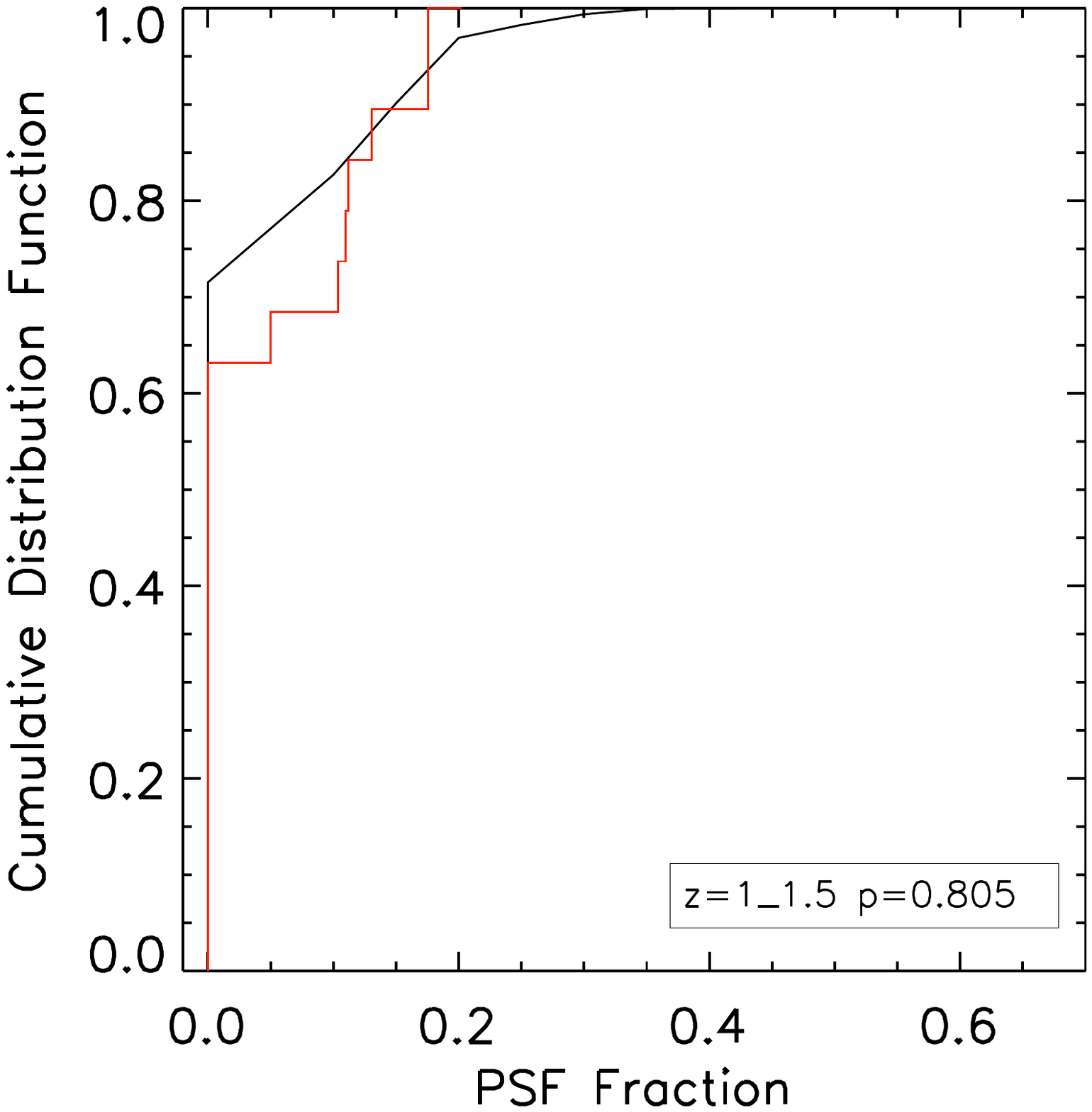}
 \includegraphics[scale=0.2, trim=0cm 13.5cm 5cm 0cm] {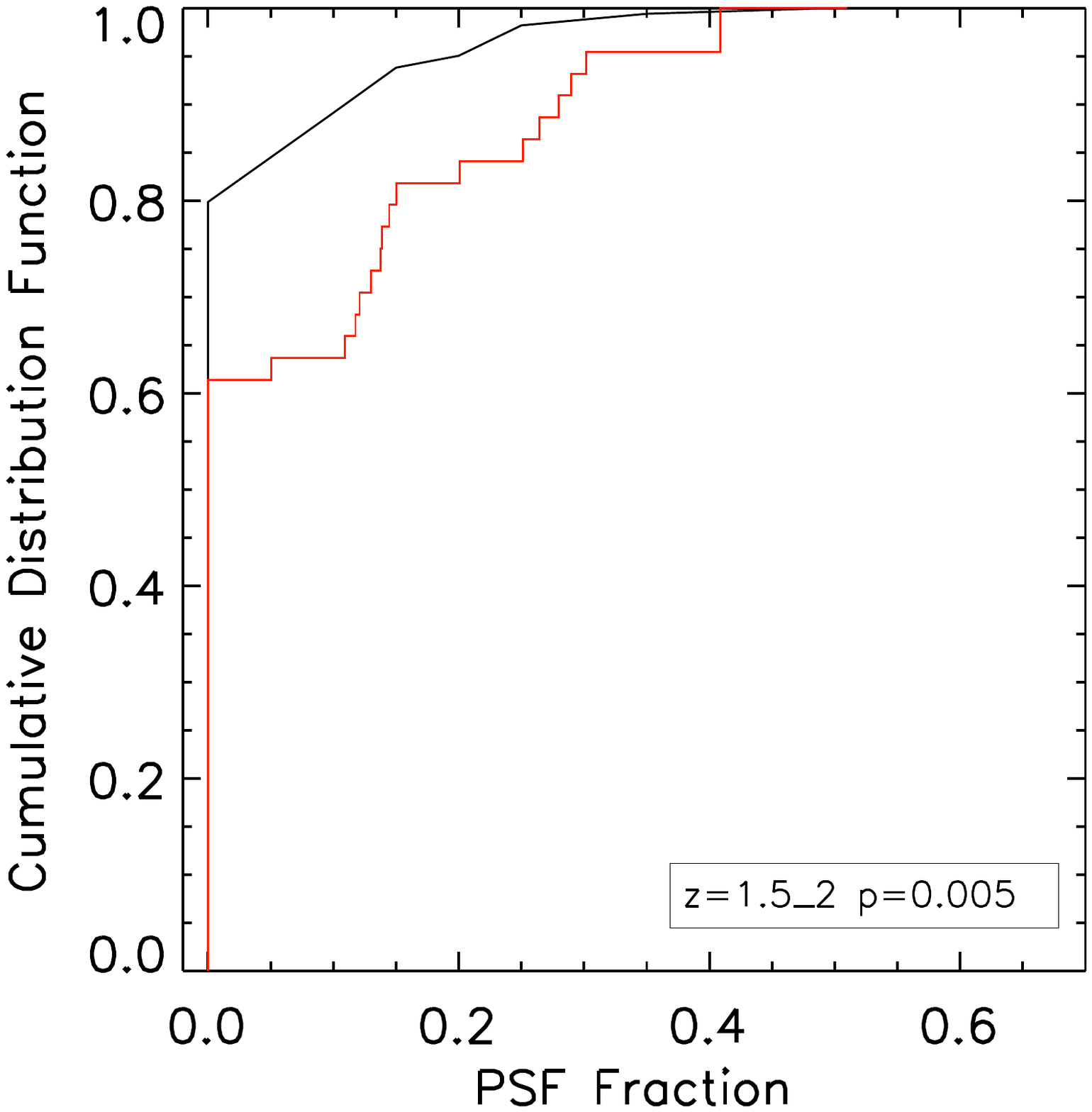}
 \includegraphics[scale=0.2, trim=0cm 13.5cm 5cm 0cm] {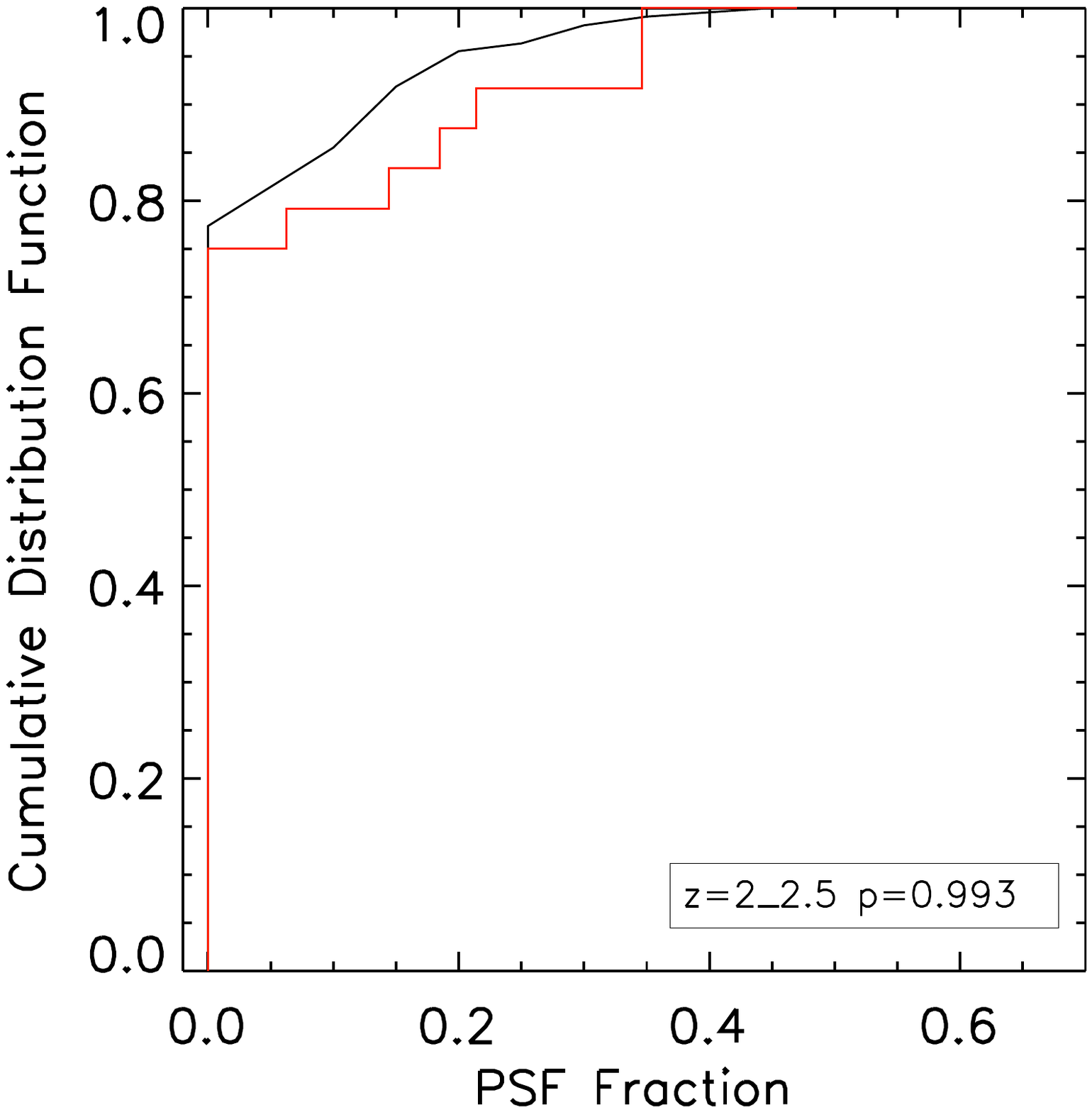}
 \includegraphics[scale=0.2, trim=0cm 13.5cm 5cm 0cm] {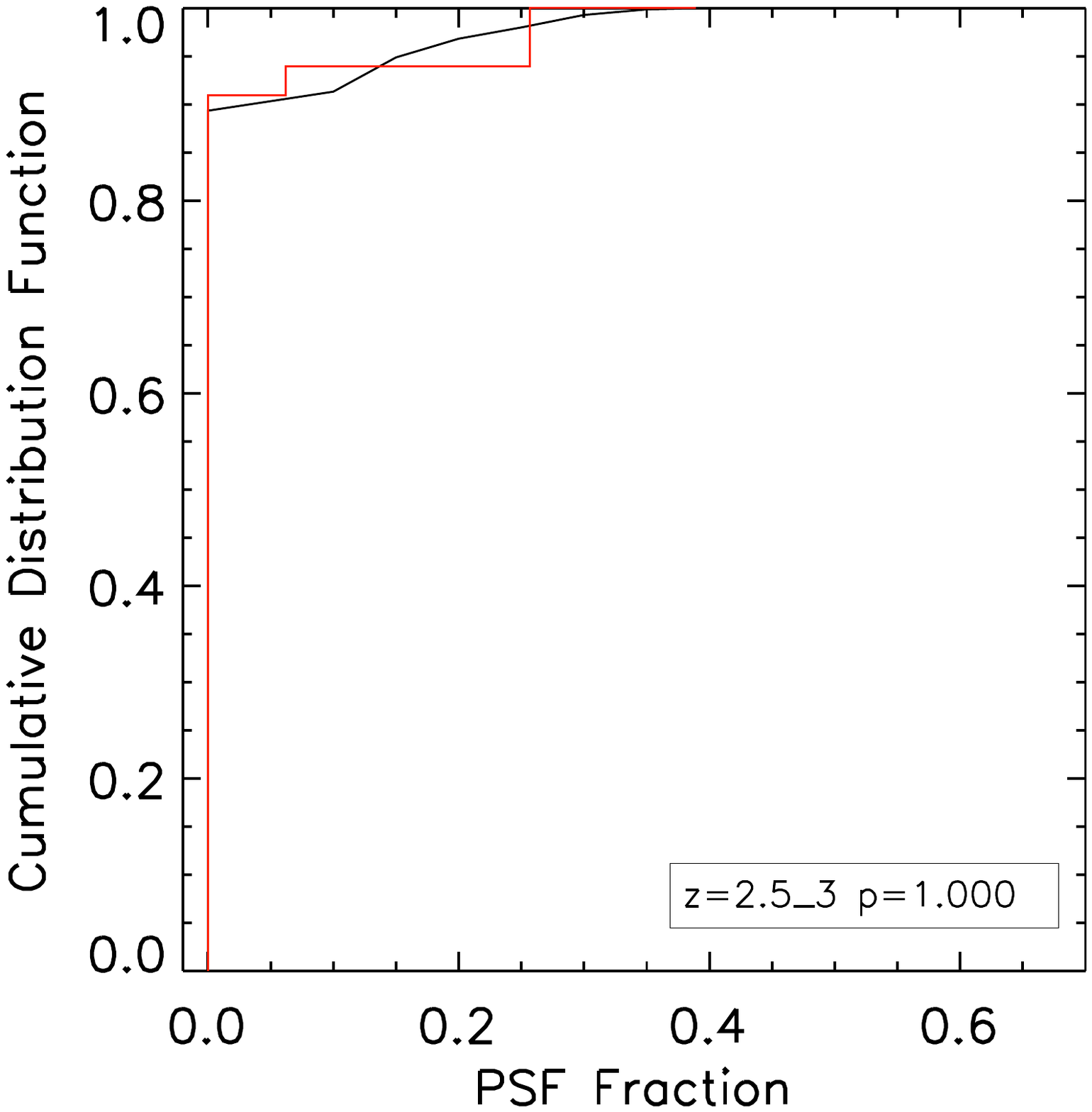}\\
 
  \includegraphics[scale=0.2, trim=0cm 13.5cm 5cm 0cm] {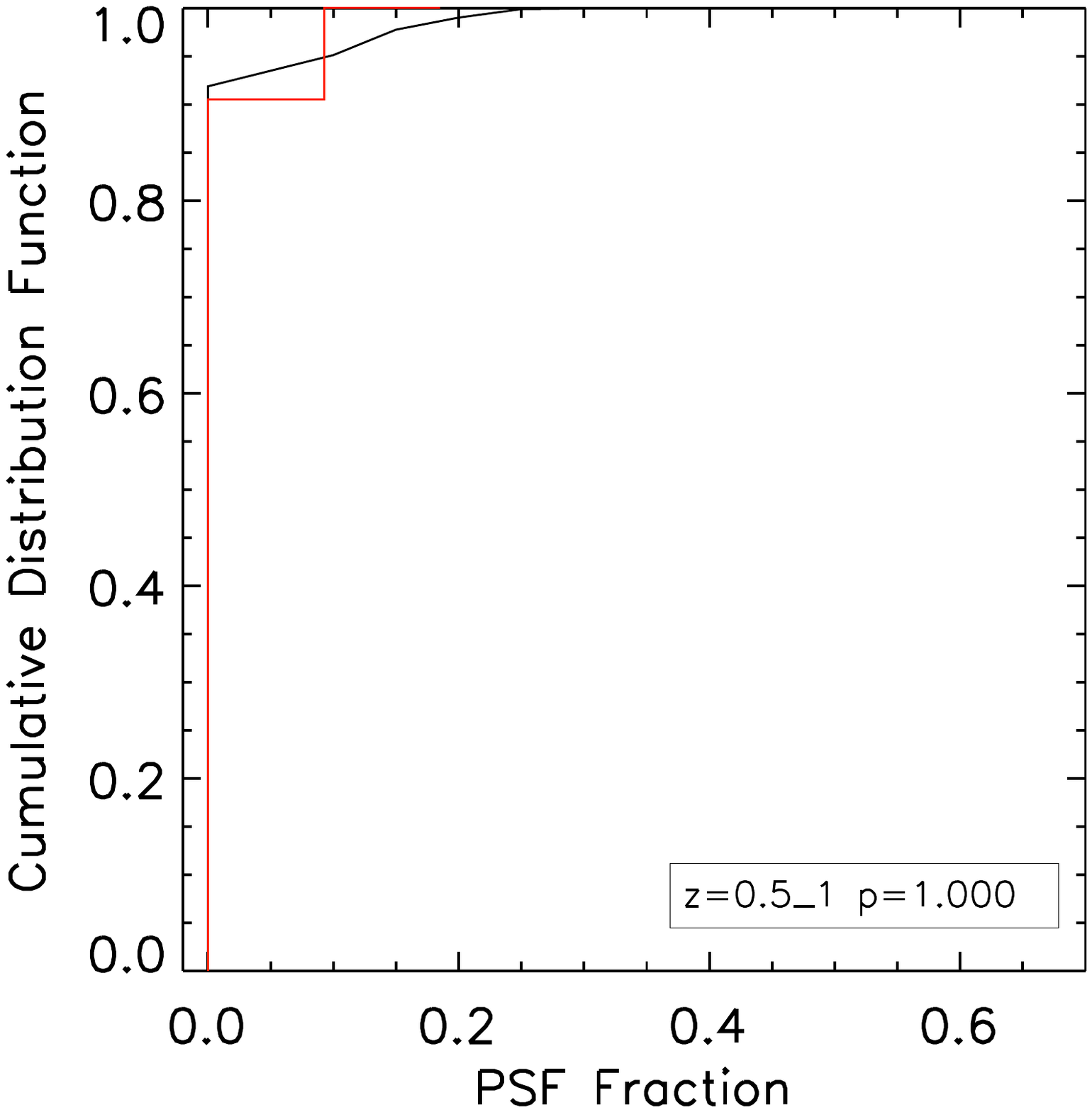}
 \includegraphics[scale=0.2, trim=0cm 13.5cm 5cm 0cm] {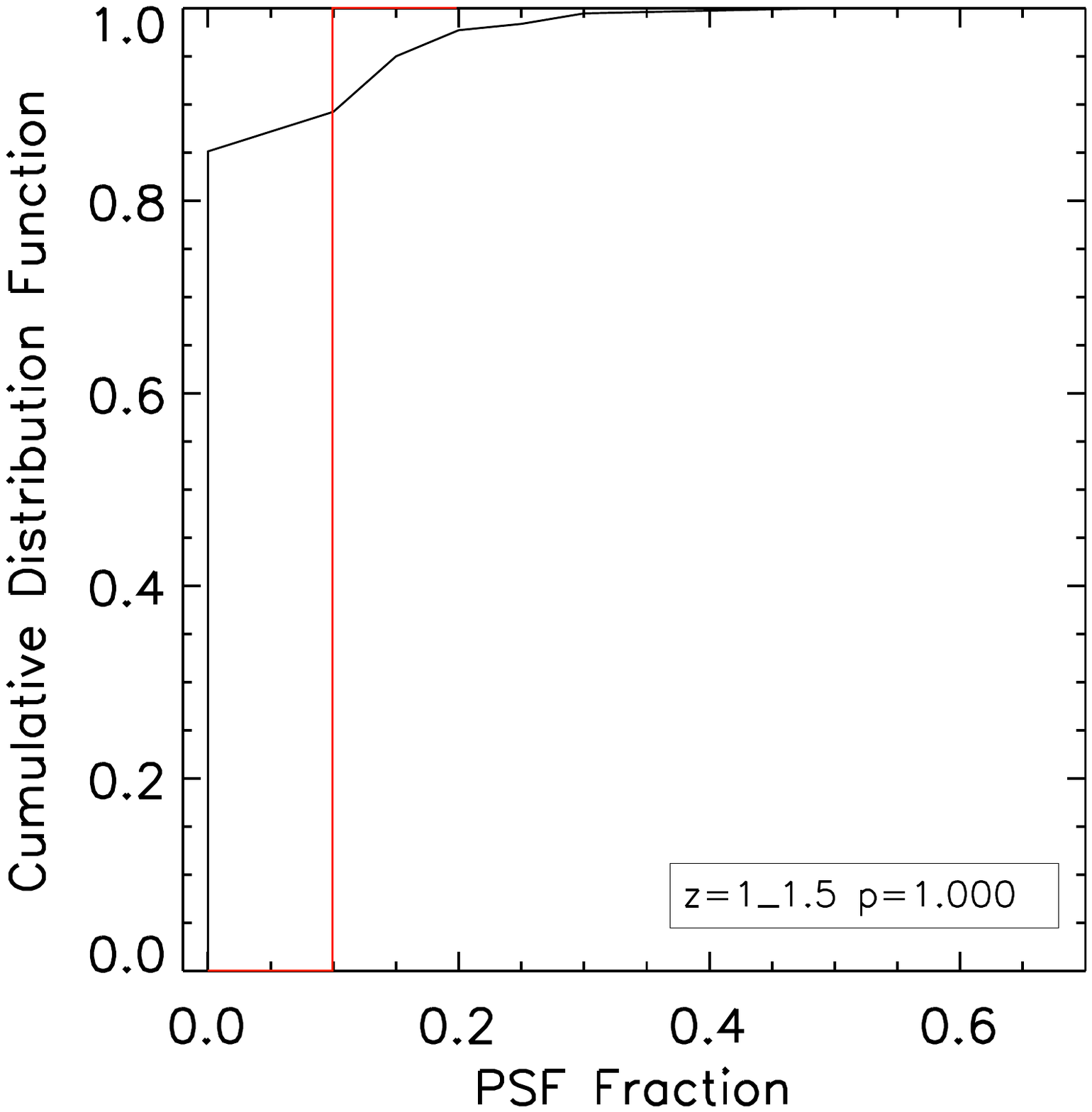}
 \includegraphics[scale=0.2, trim=0cm 13.5cm 5cm 0cm] {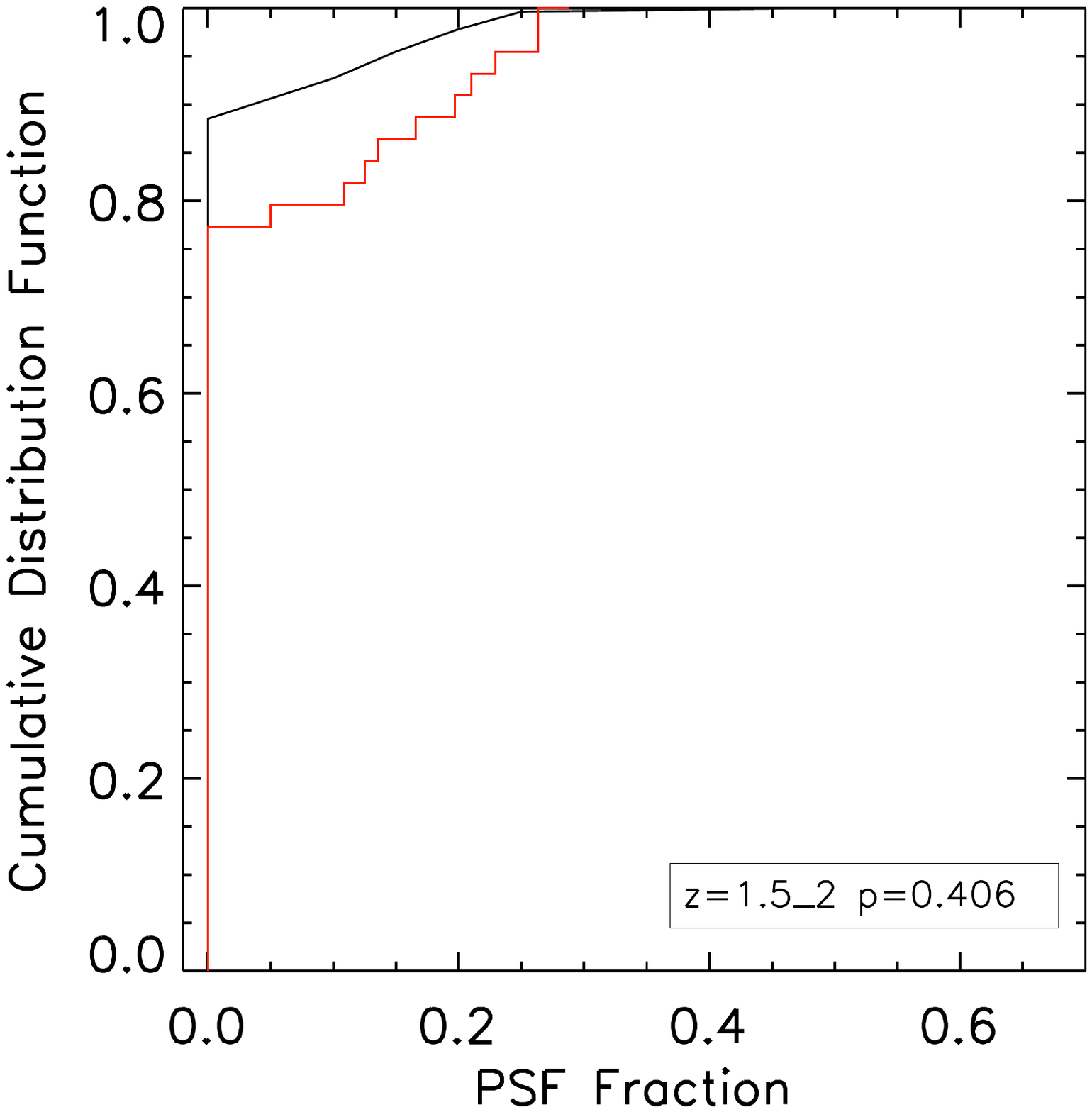}
 \includegraphics[scale=0.2, trim=0cm 13.5cm 5cm 0cm] {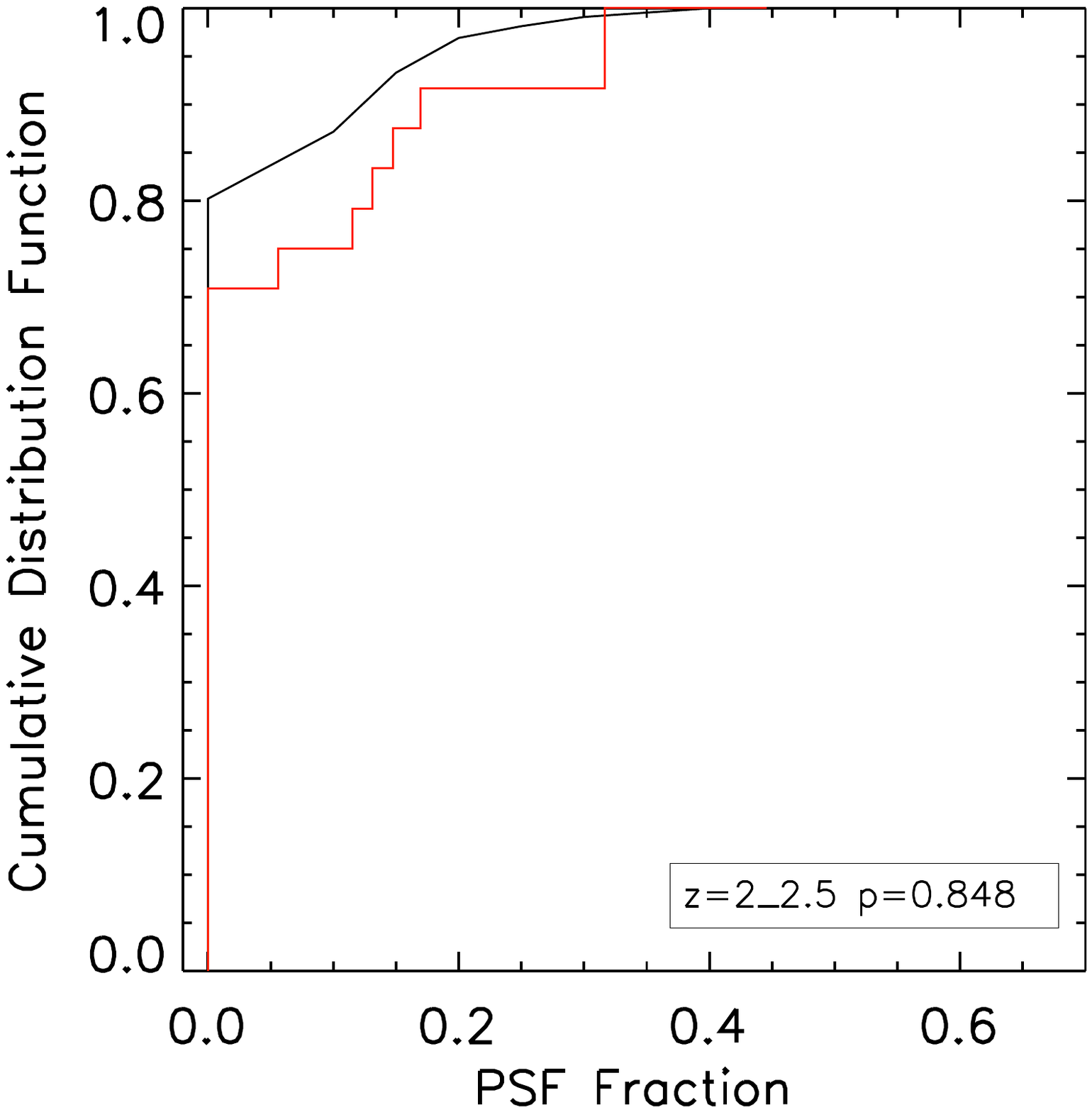}
 \includegraphics[scale=0.2, trim=0cm 13.5cm 5cm 0cm] {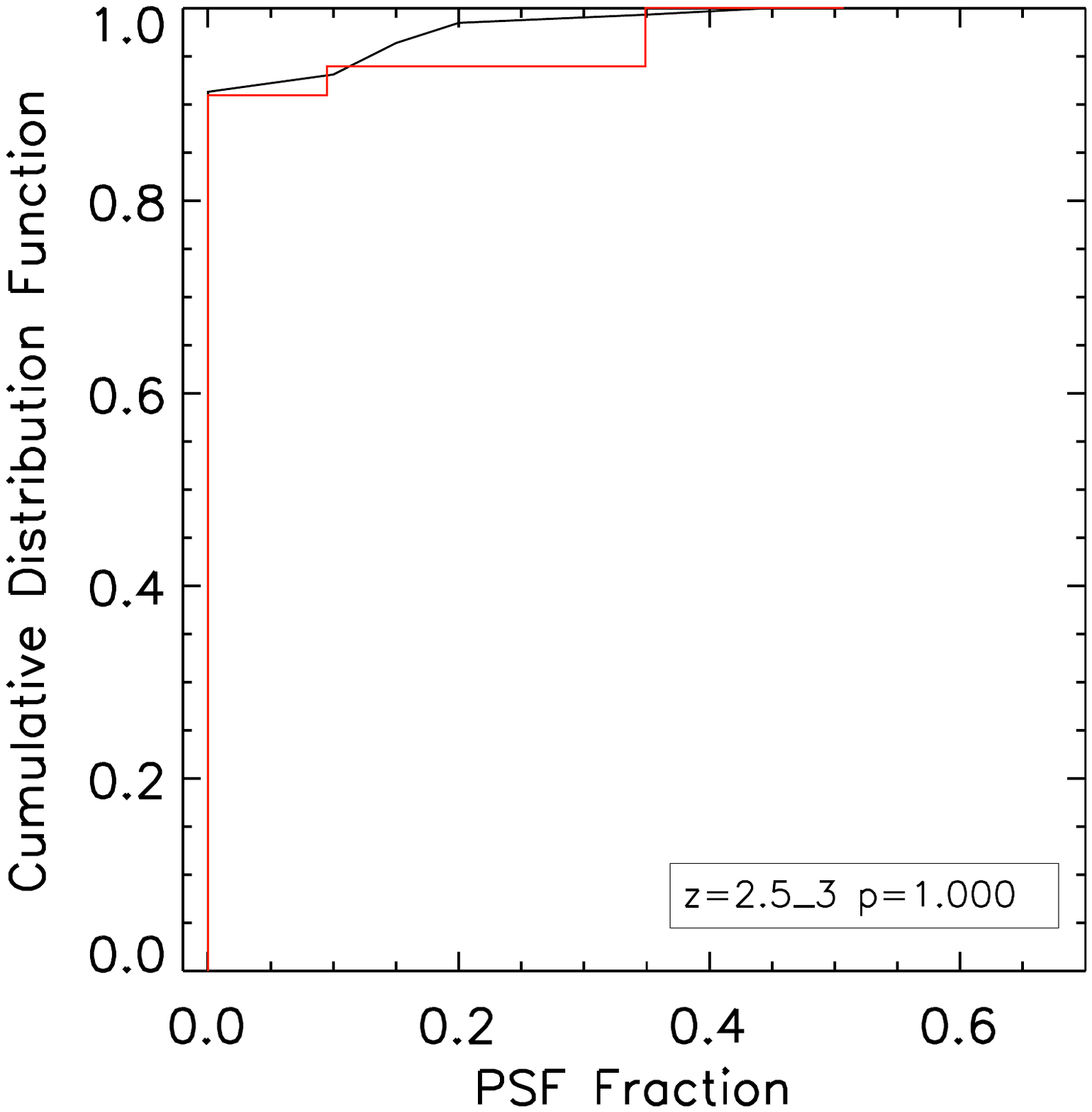}\\ 

  \caption{This figure demonstrates the similarity in the point-source light fractions between the AGN hosts and the control sample for both the single S\'{e}rsic+point-source fits and the fully decomposed bulge+disk+point-source fits. As with the previous plots, the cumulative distribution functions for the AGN are shown in red and those for the control sample are plotted in black. The point-source light fractions from the single S\'{e}rsic+point-source fits are plotted in the top row, and the fits from the full morphological decomposition are given in the bottom row. These plots serve to illustrate that there is no significant difference in the point-source light fractions between the active and non-active samples, thus suggesting that there is no strong correlation between the presence of a point-source fit in the data with X-ray selected AGN activity.}
 
 \end{figure*}

 We explore this result further by splitting our X-ray selected AGN hosts into obscured and unobscured sub-sets using both the X-ray spectra, implementing hardness ratio cuts and the effective photon index cuts from \citet{Xue2011}, and alternatively via optical spectra classifications from \citet{Szokoly2004}. In doing so we find a marginal difference in the point-source fractions for the X-ray defined obscured and unobscured sub-sets of p=0.07. As shown in Fig. 11, the unobscured AGN appear to have more low point-source fraction fits than the obscured population fits, perhaps somewhat counter-intuitively.
 Splitting by optical classifications does not reveal any difference between the point-source fractions of the obscured and unobscured subsets (p=0.96).
Looking further into the trends between the optical classifications of the X-ray selected AGN and the prevalence of a point-source fit in their morphologies, we observe no clear trend between the brightest QSOs and point-source fractions. However, (as can be seen in Fig. 12) by splitting the AGN sample at $L_{X}=1\times 10^{43.5} {\rm erg\,s^{-1}}$, the limit above which traditionally AGN contamination is thought to become a concern at X-ray luminosities, we do find some evidence that the most luminous AGN have higher point-source light fractions, where this result is significant at the $\simeq99.7\%$ confidence level from a  K-S test.

As an aside, on an object-by-object basis, $9.7\pm0.8\%$ of all galaxies are best fit with a point-source component compared to $13.4\pm2.4\%$ of all X-ray selected AGN hosts. This suggests a marginal preference for AGN hosts to be more point-source dominated but as mentioned previously there are mass biases implicit in this comparison which make it less than ideal.

From these comparisons we see no definitive confirmation that the point-source fits correlate with either the presence of an X-ray selected AGN, or within the AGN sample with the AGN obscuration, as might be expected. 

   \begin{figure}
 \centering
  \includegraphics[scale=0.28, trim=4.5cm 12cm 8cm 0cm] {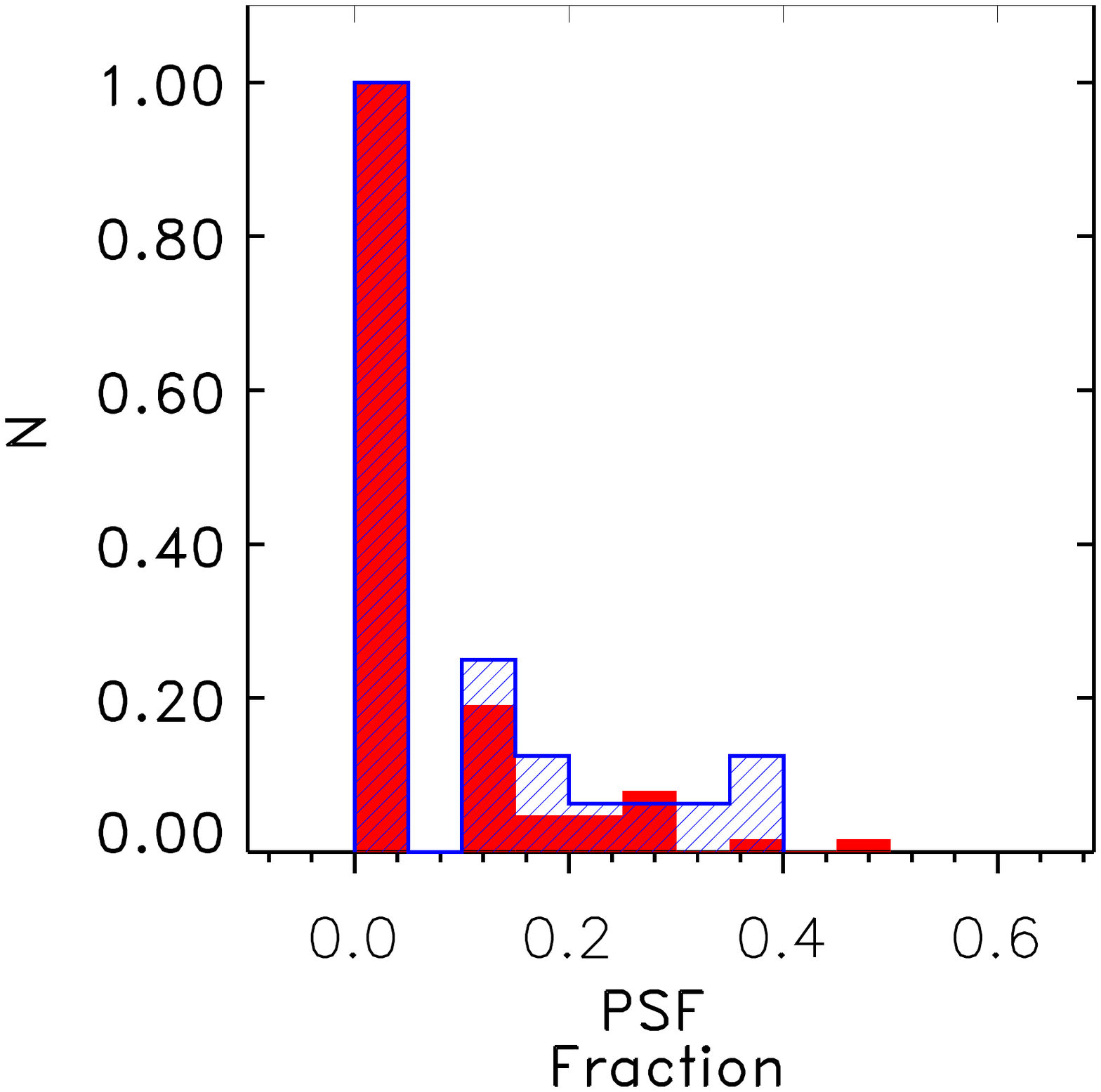}
  \caption{The point-source light fractions of the X-ray selected AGN sample split into low X-ray luminosity, $L_{X}<1\times 10^{43.5} {\rm erg\,s^{-1}}$ (red) and high luminosity $L_{X}>1\times 10^{43.5} {\rm erg\,s^{-1}}$ (blue) using samples. These distributions have been normalised such that the maximum bin value is set equal to one. This plot reveals that the higher X-ray luminosity AGN have higher point-source light fractions compared to the lower luminosity sample.}
 
 \end{figure}
  \begin{figure}
 \centering
  \includegraphics[scale=0.28, trim=4.5cm 12cm 4cm 0cm] {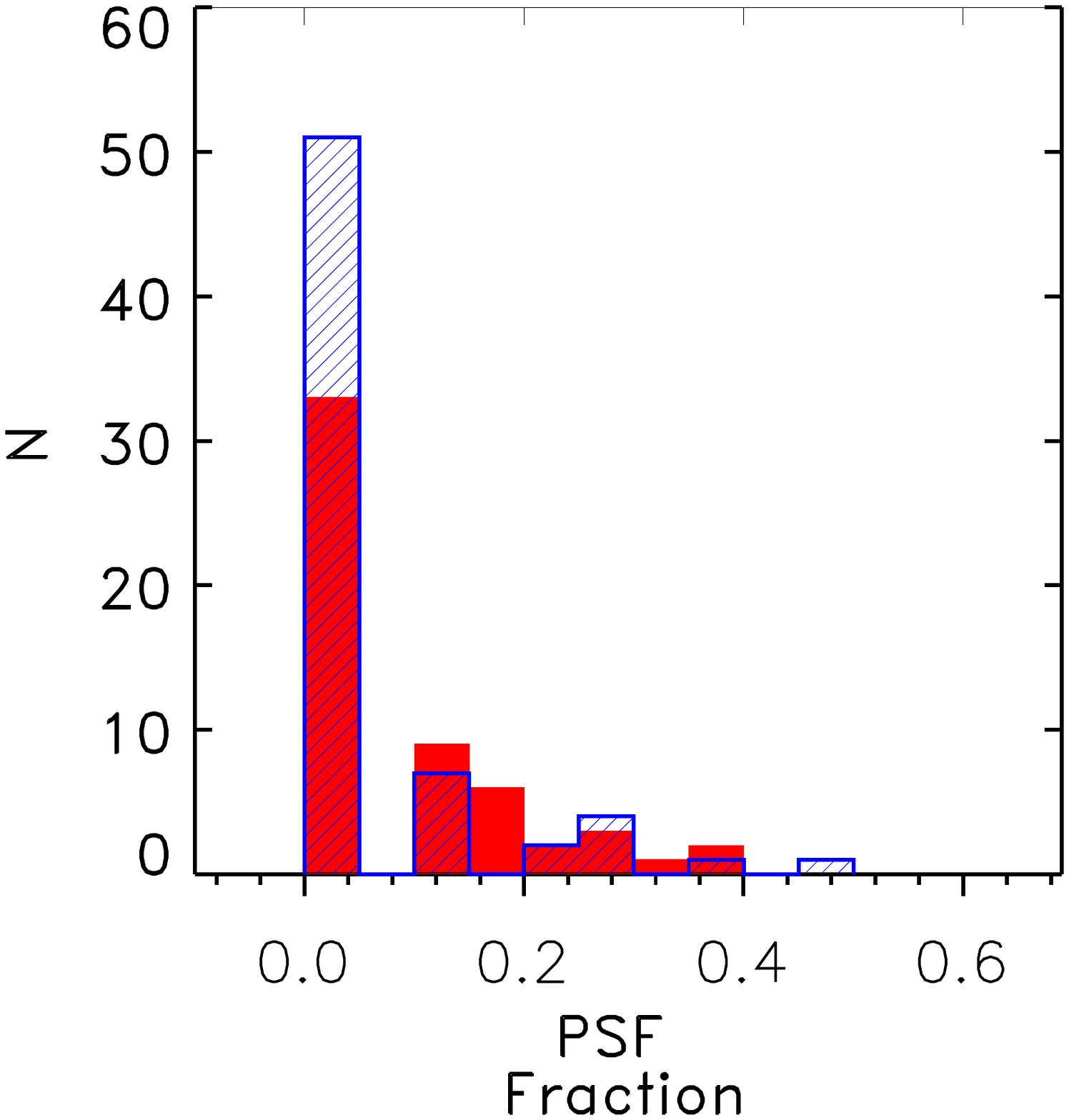}
  \hspace{0.3cm}
   \includegraphics[scale=0.28, trim=2.0cm 12cm 4cm 0cm] {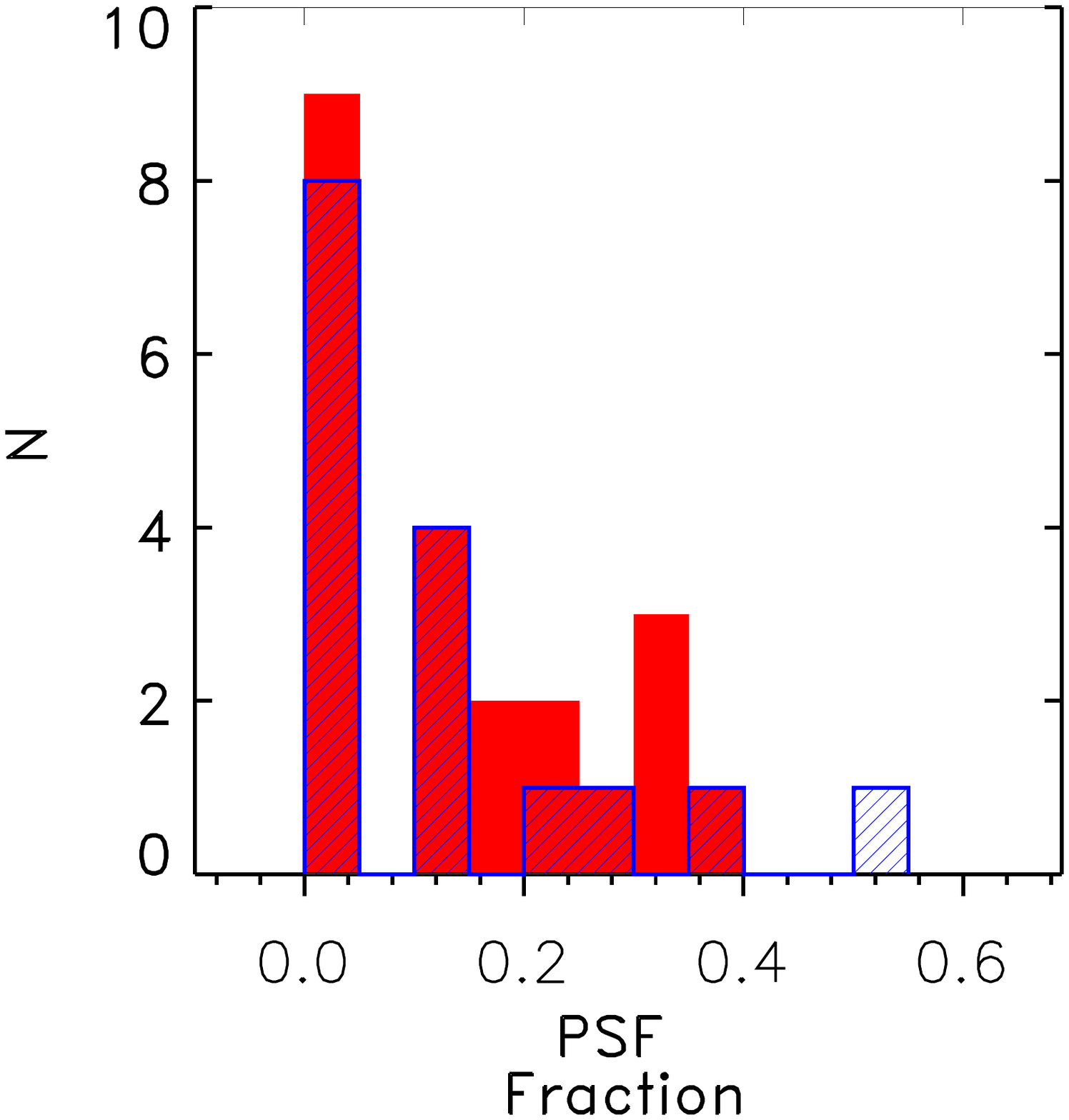}
  \caption{Here we plot the point-source light fractions of the X-ray selected AGN sample split into obscured (red) and unobscured (blue) using : left, hardness ratios and effective photon index parameters from \citet{Xue2011}; right, optical spectra classifications from \citet{Szokoly2004}. These distributions hint at  a puzzling trend for the X-ray classified unobscured AGN to have lower point-source fractions than the obscured population.}
 
 \end{figure}

  \subsection{Stacking of X-ray Images}
  In order to ensure this result is not biased by point-source fits which are AGN with faint X-ray signatures below the detection limit of the \citet{Xue2011} catalogues, we have constructed stacks of the X-ray image at the CANDELS positions of the sources with: the 10 brightest point-source components; and with the 10 largest point-source fractions. For direct comparison we also stack 10 non-point source fits with no X-ray detections in the catalogue with similar redshifts, masses, magnitudes and sizes as the point-source sample in order to construct a control set. The stacks have been created from a mean combination.
  
  The stacks are displayed in Fig. 13 with the top row for the brightest point-source fits and the bottom row for the  point-source fits with the highest light fractions. The stamps on the left are the mean stacks of the point-source fits with no X-ray detections and the stamps on the left illustrate the control samples for both cases. Aperture photometry on these stacks reveals no significant evidence for more flux in the stacks of the highest point-source light fractions sample than in the stacks of the control sample.

 \begin{figure*}
 \centering

  \includegraphics[scale=0.4, trim=3cm 13cm 8cm 5cm] {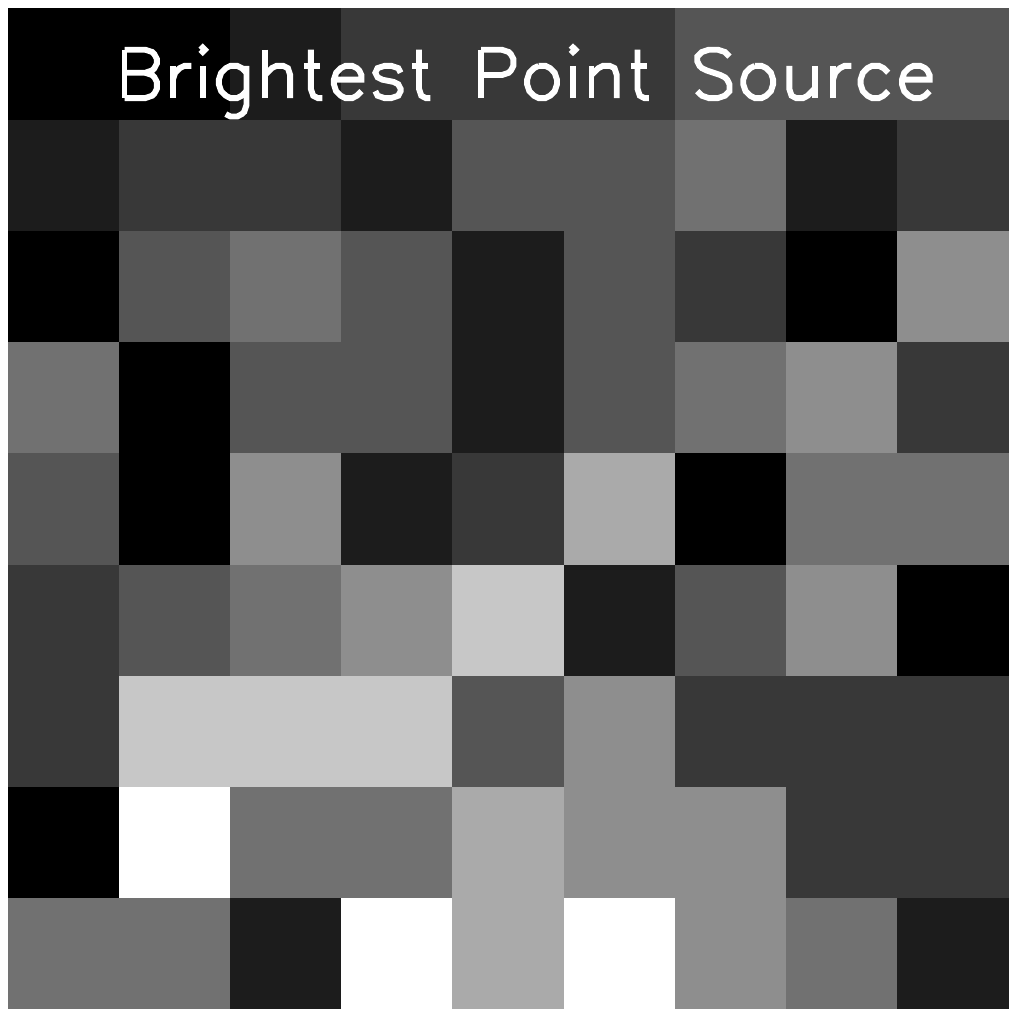}
  \hspace{0.2cm}
 \includegraphics[scale=0.4, trim=3cm 13cm 8cm 5cm] {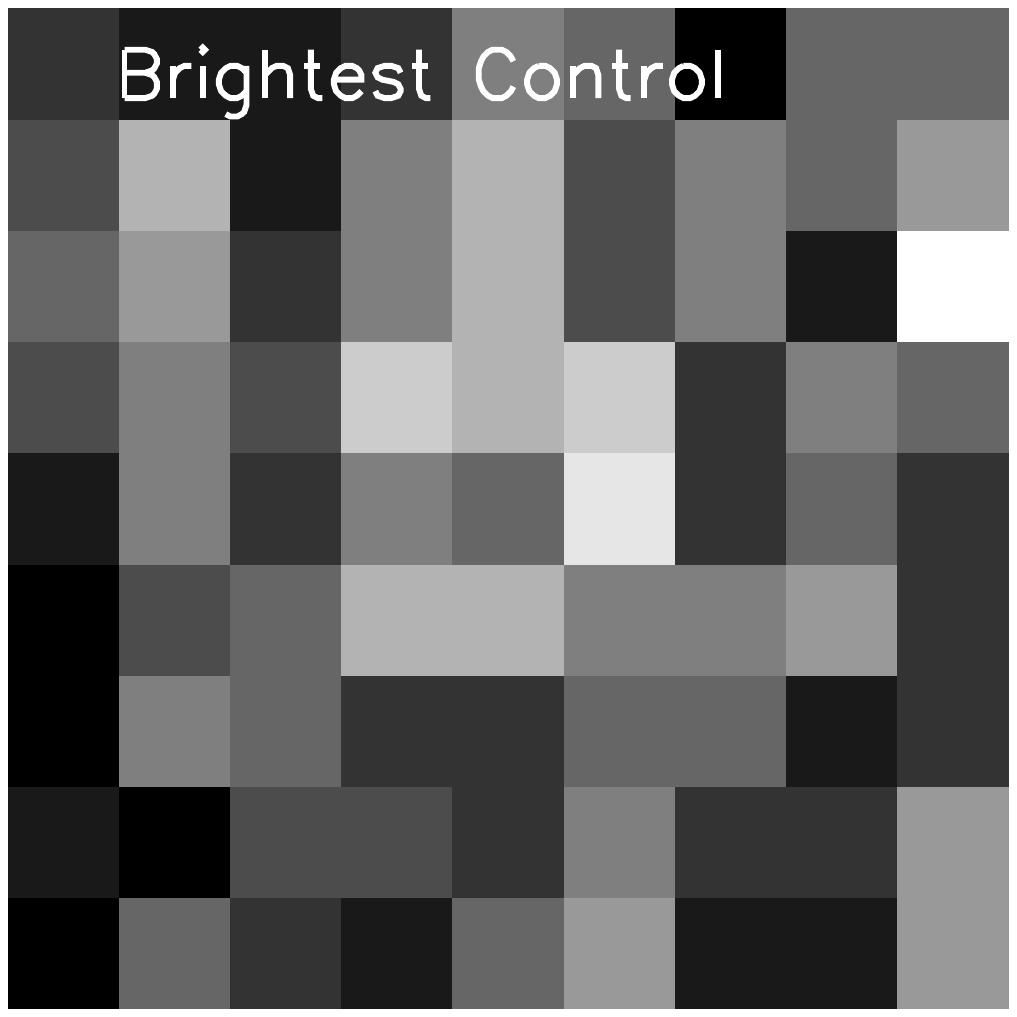}
   \hspace{0.2cm}
\includegraphics[scale=0.4, trim=3cm 13cm 8cm 5cm] {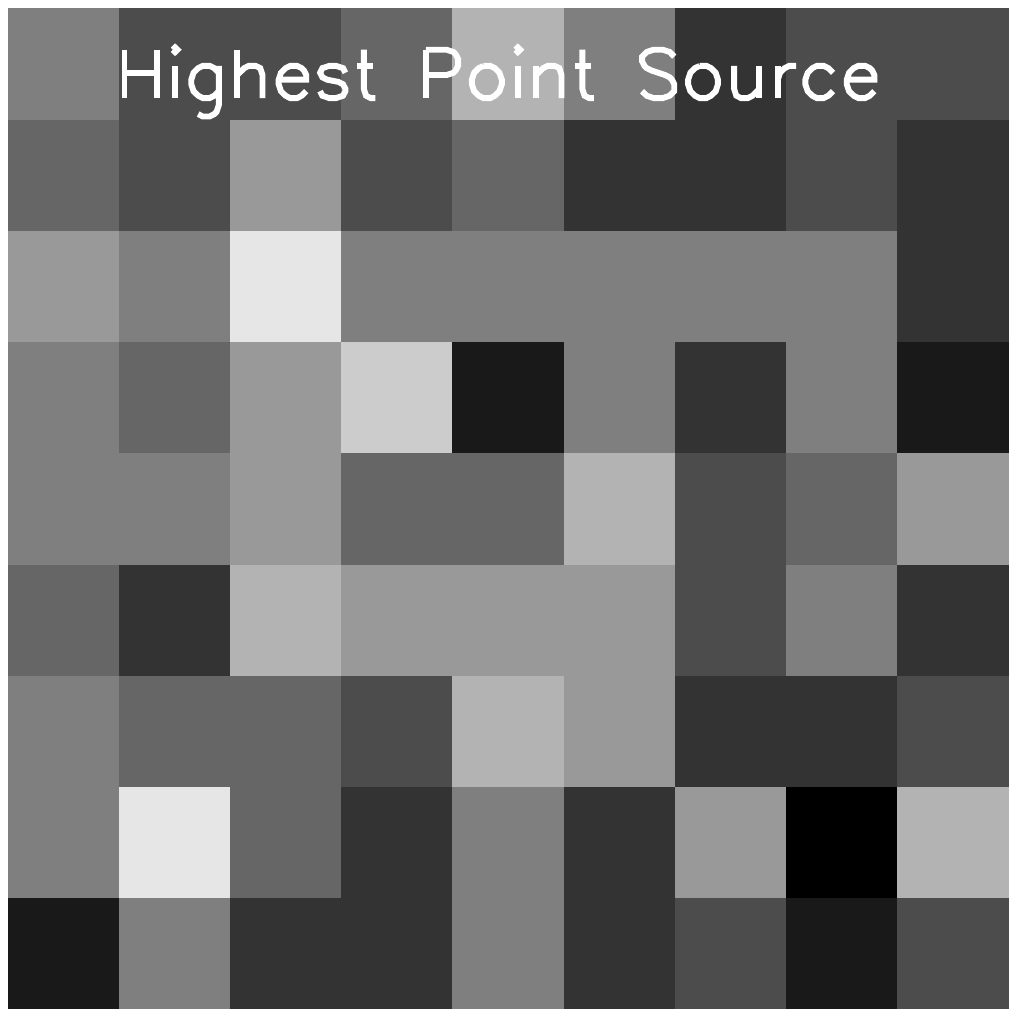}
  \hspace{0.2cm}
 \includegraphics[scale=0.4, trim=3cm 13cm 8cm 5cm] {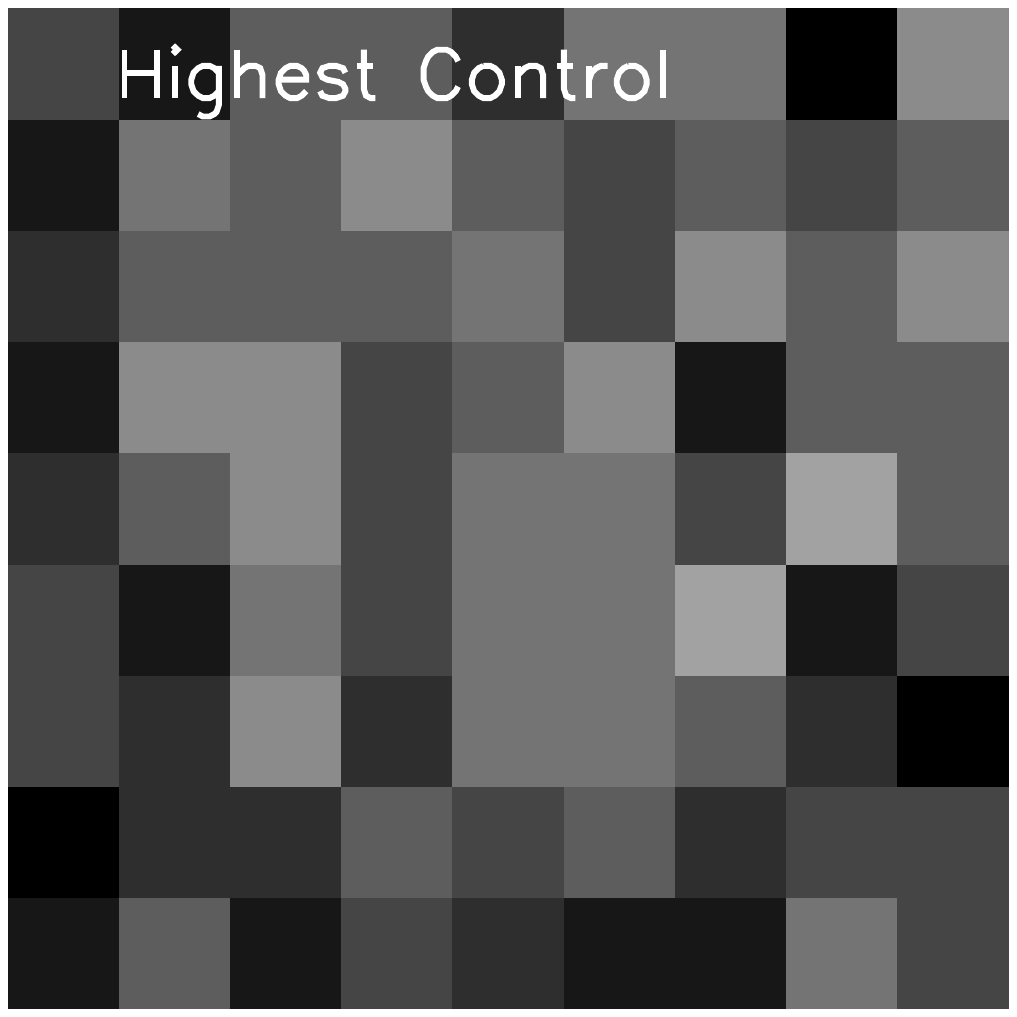}\\ 
  \caption{The $\sim 2\times 2$ arcsec X-ray stacked images of the various sub-samples probed in order to explore the possibility of point-source model fits picking up faint AGN below the detection limits of the \citet{Xue2011} X-ray selected catalogue. From left to right are: the mean combined stacks of the 10 brightest PSF fits without X-ray counterparts and the associated non-point source control sample, then the stacks of the 10 highest fraction point-source fits without X-ray detections and the control sample. Here we can see no evidence for excess flux in the stacks of either of the point-source model samples and as a result conclude that the point-source fits are not indicative of AGN activity in the X-ray below the limit of current surveys. }
 
 \end{figure*}

 \subsection{Further Connections Between Point Source and X-ray Properties}
 
 In the previous section we have concluded that, by comparing the point-source fractions between AGN hosts and a mass-matched control sample there is no evidence for a correlation between the prevalence of a point-source fit and an AGN detection, even when stacking the X-ray to search for low luminosity sources. However, another correlation to test for is the link between the point-source luminosity and the luminosity of the AGN detected in the X-ray, which should shed light on whether the point-source fit is related to AGN activity (in which case one would expect a scaling between point-source luminosity and AGN X-ray luminosity) or whether it is a signature of a nuclear starburst in the system.   
\begin{figure}
 \centering
 \includegraphics[scale=0.5, trim=0cm 13cm 0cm 0cm]{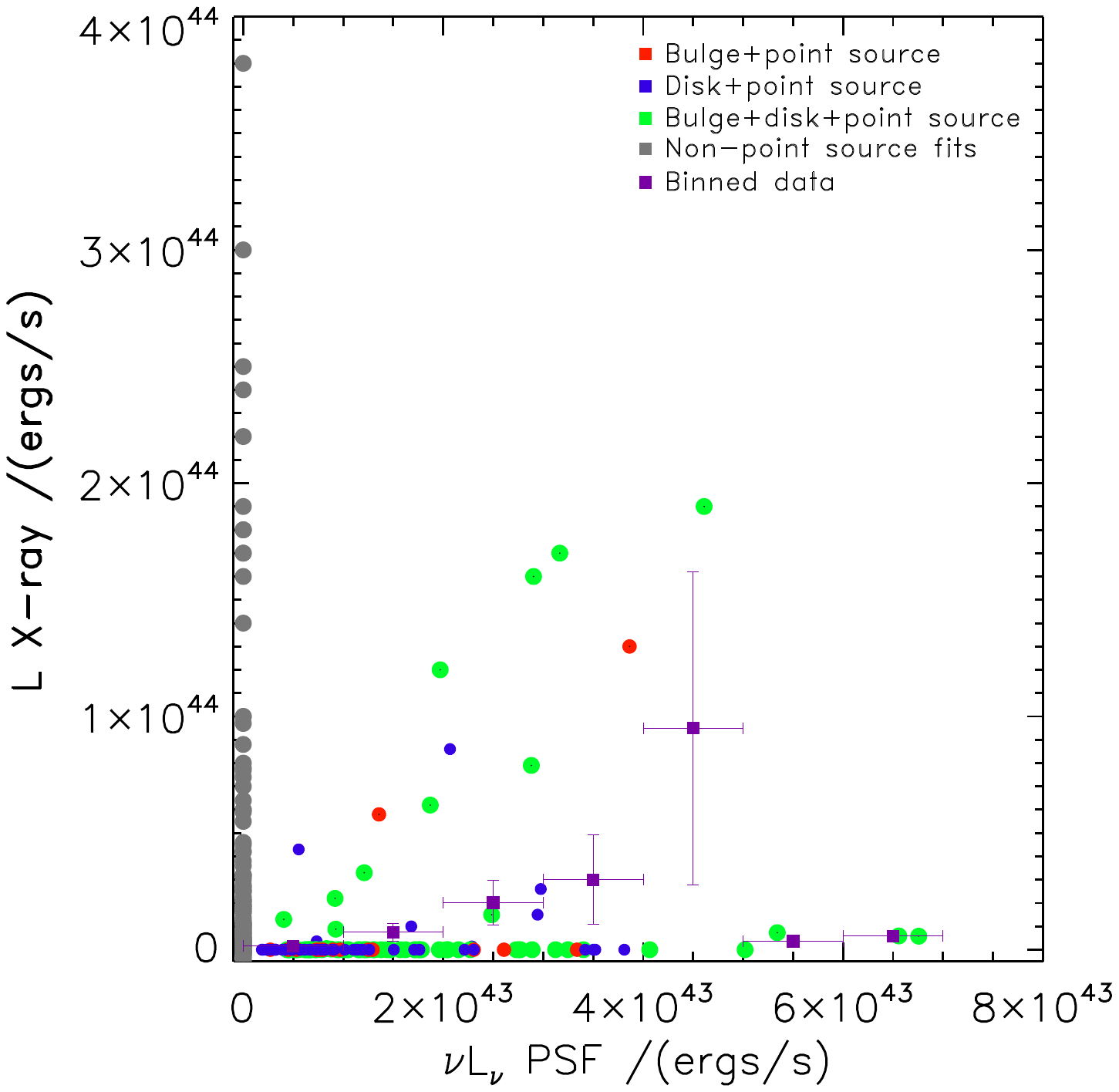}
 \caption{This figure explores the potential correlation between the AGN X-ray luminosity and the point-source luminosity in the H$_{160}$ band. Objects which have a point-source fit with or without an X-ray counterpart are colour coded: red for bulge+point-source fits, blue for disk+point-source fits and green for bulge+disk+point-source fits. Objects which have an X-ray counterpart with no point-source fit are coloured grey. We also over-plot in purple the binned mean of the X-ray luminosity in each point-source luminosity bin with error-bars representing the standard error on the mean.}
 \end{figure}
 
This comparison is given in Fig. 14, now on an object-by-object basis. Here we plot the objects with X-ray detections and point-source component fits, as well as point-source fits with no X-ray counterparts in colour, where red represents bulge+PSF fits, blue represents disk+PSF fits and bulge+disk+PSF fits are plotted in green. Objects with an X-ray detection and no point-source fits are plotted in grey. There is significant scatter in this plot but there is a positive correlation between the point-source luminosity and the AGN X-ray luminosity, which can be fitted with a Spearman Rank correlation coefficient $\rho=0.31$ with $p<0.005$. When binned, as shown in purple, we find that the correlation between the mean values in each bin is weaker and no longer statistically significant: $\rho=0.143$ with $p>0.25$.

  \begin{figure}
 \centering
  \includegraphics[scale=0.35, trim=4cm 13cm 0cm 0cm] {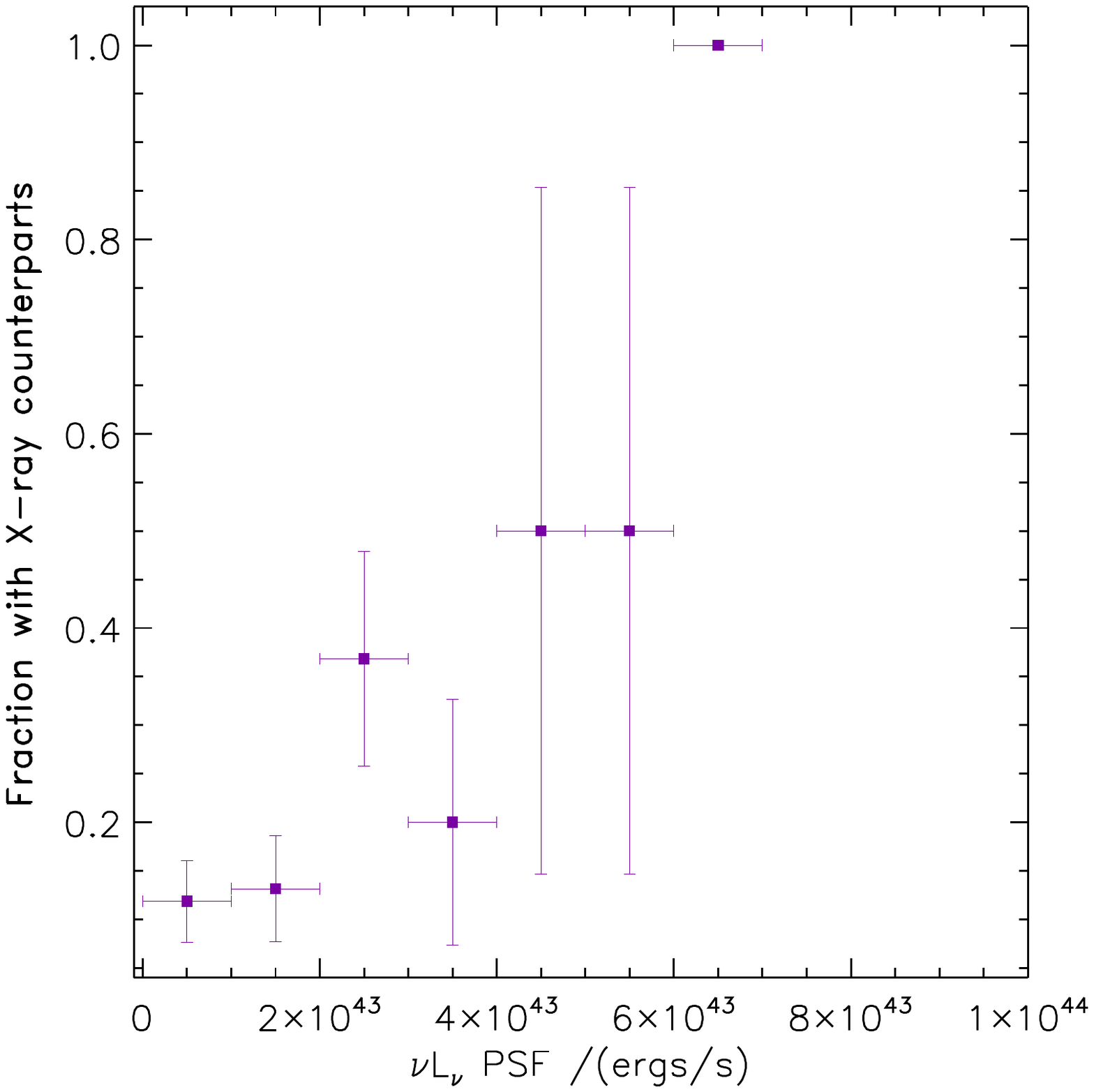}
\caption{Plot demonstrating the fraction of point-source fits with an X-ray counterpart as a function of the point-source component luminosity. These results suggest that the most luminous point-source fits are the most likely to host AGN. }
 \end{figure}
 
To explore this potential connection further we have also looked at the fractions of point-source components which have X-ray counterparts (Fig. 15). Whilst there is tentative evidence here that the fraction of point-source components with X-ray counterparts increases with the luminosity of the point source, the errors on these fractions are large as they are driven by the scatter within the sample.

  \begin{figure}
 \centering
  \includegraphics[scale=0.35, trim=4cm 13cm 0cm 0cm] {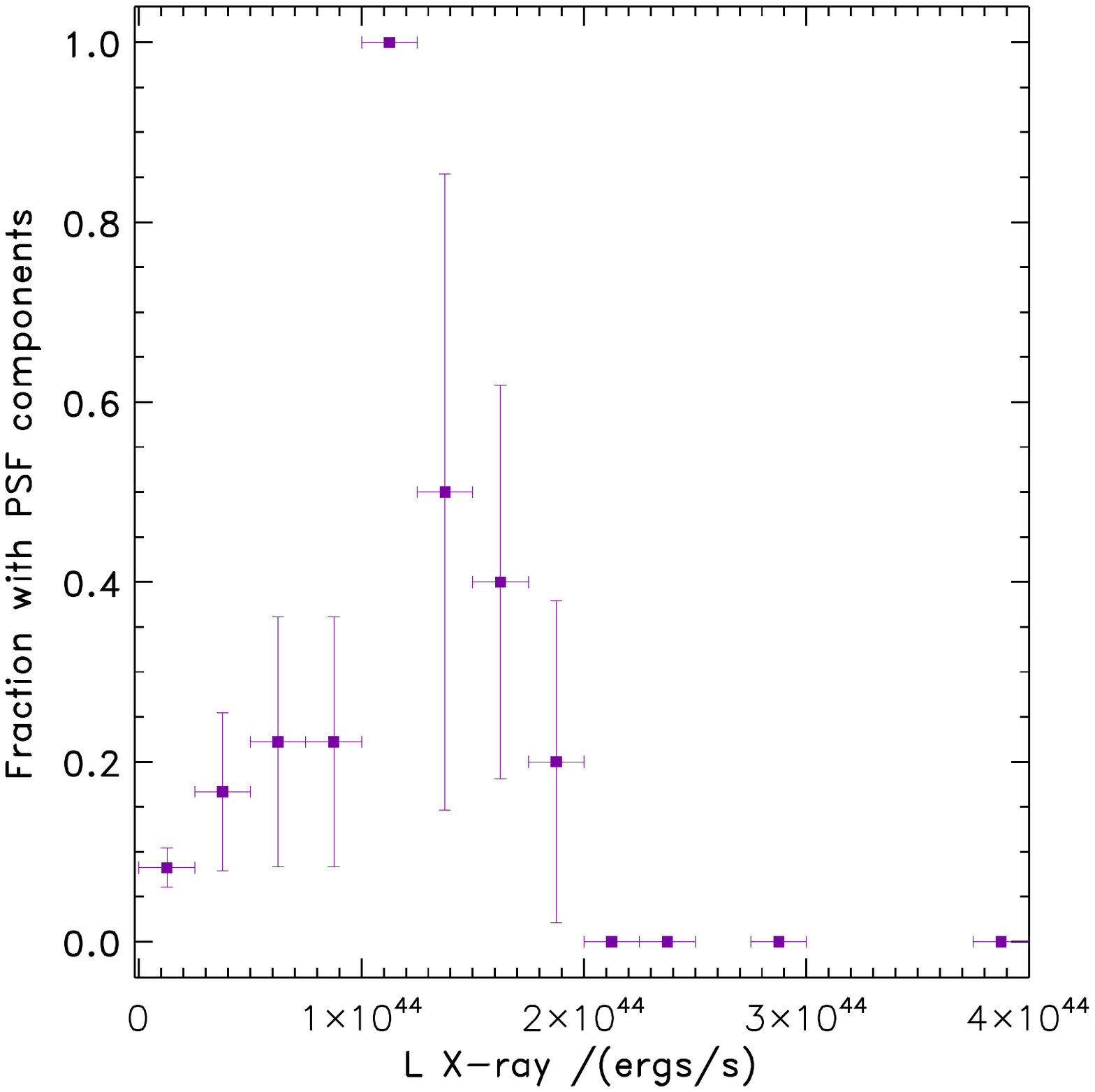}
\caption{ In contrast to Fig. 15, here we present the fraction of X-ray counterparts with point-source fits as a function of the X-ray luminosity. Now we see a turnover at moderate luminosity: below which there appears to be a general trend toward more luminous AGN having an increased likelihood of point-source component fits; and above which the opposite is the case.}
 \end{figure}
 
 Conversely, it would seem that the fraction of X-ray sources with a point-source component fit (Fig. 16) correlates with the moderate luminosity AGN but then falls off at higher AGN X-ray luminosities.  Moreover, this trend is unaffected if the AGN sample is split according to effective photon index into obscured and unobscured sub-samples.

 In conclusion there does not appear to be a strong correlation between the adoption of a point-source component in the morphological fitting and the presence of an AGN. However, there is some indication that by selecting the brightest point-source component fits, one is more likely to be preferentially selecting AGN hosts, but this bright point-source fit sub-set may also contain non-active galaxies. This lack of a strong correspondence between point-source fits and AGN activity is not surprising given that as we show in a companion paper (Bruce et al., in preparation) that the majority of point-source components are best fit by a stellar component and so are best physically modelled by a nuclear starburst.

 \section{Conclusions}
 We have presented an analysis of the rest frame optical morphologies of AGN hosts in the CANDELS GOODS-S field in the redshift range $0.5<z<3$ with masses $M_*>10^{10}{\rm M_{\odot}}$ and compared them to a carefully mass-matched control sample of non-active galaxies. Having explored the effects of fitting either single S\'{e}rsic, multiple S\'{e}rsic and models with point-source components, we confirm that the AGN hosts are indeed more bulge dominated in the highest redshift bins ($z\geq2$) than the non-active control sample, as has been reported from previous studies. We conclude that morphological fits of AGN without the addition of point-source components to model any nuclear light from the AGN do not bias this overall trend towards AGN hosts being more bulge dominated. Furthermore, when fully decomposed into their separate components it is clear that the AGN hosts have significant bulge and disk components. 
 
 This is suggestive of the fact that both of these stellar populations play an important role in the triggering of AGN, with not only the bulge component being crucial to feed the central black hole, but a massive disk being necessary to support the growth of the bulge.

 Our results appear to be in agreement with a more general shift in the literature away from the scenario where merger driven accretion is thought to be the dominant fuelling mode at $z\geq1$ for moderate luminosity AGN. By demonstrating that the AGN hosts display massive disk components and that the overall evolution of the AGN hosts across the $0.5<z<3$ range is comparable with that, although may be accelerated in comparison to, the non-active control galaxies, our findings also point towards a more secular process being responsible for the triggering of AGN.
 
 From a full exploration of the correlation between point-source fits in a purely mass-selected sample of galaxies and the prevalence of AGN activity within these sources, we do not find strong evidence in favour of the point sources being AGN in nature. In fact, there is considerable evidence from the colours of the point-source fits, as will be shown in a companion paper (with a full exploration of the trends between the fitted ages, dust obscuration values and morphologies), that they are in fact nuclear starbursts. Consequently, the adoption of the point-source fits in both the active and non-active control samples should be used to best describe the full stellar structure of these systems.
 
 Finally, from a comparison of the total stellar mass and bulge stellar mass trends within our full mass-selected and AGN host samples, we find evidence in this redshift regime that the BH mass may be better correlated with the bulge mass than the total stellar mass of these systems. Furthermore, we also find evidence that the trend for AGN hosts to have a higher low-end cut-off in bulge stellar mass than the underlying galaxy population may play a role in the more general observation that AGN hosts are biased towards higher total stellar masses compared to the underlying galaxy population within a mass-cut sample. 
  
\section{Acknowledgements}

VAB and JSD acknowledge the support of the EC FP7 Space project ASTRODEEP (Ref. No: 312725).
JSD acknowledges the support of the 
European Research Council via the award of an Advanced Grant. 
AM acknowledges funding from the STFC and a European Research Council Consolidator Grant (P.I. R. McLure).

This work is based in part on observations made with the NASA/ESA {\it Hubble Space Telescope}, which is operated by the Association 
of Universities for Research in Astronomy, Inc, under NASA contract NAS5-26555.
This work is based in part on observations made with the {\it Spitzer Space Telescope}, which is operated by the Jet Propulsion Laboratory, 
California Institute of Technology under NASA contract 1407.

\bibliographystyle{mn2e}

\bibliography{bibtex}

\label{lastpage}
 
\end {document}